\documentclass{jaa}
\usepackage{natbib}
\usepackage{layout}
\usepackage[utf8]{inputenc}
\usepackage{import}
\usepackage{multicol}
\usepackage{amsmath}  
\usepackage{amssymb}  
\usepackage{ulem}
\usepackage{placeins}
\usepackage{epsfig}
\usepackage{cuted}
\usepackage{multirow} 
\usepackage{booktabs} 
\usepackage{float}    
\usepackage{stfloats}
\usepackage{adjustbox} 
\usepackage[table,xcdraw]{xcolor}
\usepackage{threeparttable}
\usepackage{bm}  
\usepackage{hyperref}
\hypersetup{
     colorlinks=true,
     linkcolor=blue,
     filecolor=magenta,
     urlcolor=cyan,
     citecolor=blue,
     anchorcolor=blue}

\usepackage{lineno}

\begin{document}\sloppy

\title{\boldmath Study of the Ultra-Long GRB 220627A: Exploring a Possible Blue Supergiant Collapsar Origin}


\author{A. Garg\textsuperscript{1,*}, A. Shukla\textsuperscript{1}, D. Sz\'{e}csi\textsuperscript{2}, G. C. Anupama\textsuperscript{3}}
\affilOne{\textsuperscript{1}Discipline of Astronomy, Astrophysics and Space Engineering, Indian Institute of Technology, Khandwa Road, Indore-453552, Madhya Pradesh, India.\\}
\affilTwo{\textsuperscript{2}Institute of Astronomy – Faculty of Physics, Astronomy and Informatics, Nicolaus Copernicus University, Jurija Gagarina 11, Toruń - 87-100, Kuyavian-Pomeranian Voivodeship, Poland.\\}
\affilThree{\textsuperscript{3}Indian Institute of Astrophysics, II Block Koramangala, Bengaluru 560034, India}


\twocolumn[{
\maketitle
\corres{ayushg827@gmail.com}

%
\begin{abstract}
Ultra-long gamma-ray bursts (ul-GRBs) challenge standard collapsar models because their extreme durations require central engines active for $\sim10^{3}$--$10^{4}$~s. While the prompt spectral evolution of typical GRBs is difficult to resolve due to their short durations, ul-GRBs provide a unique opportunity to study spectral evolution on extended timescales. We present a time-resolved spectroscopic study of GRB~220627A, one of the most luminous ul-GRBs known ($E_{\rm iso}\sim10^{54}$~erg, $T_{90}>1000$~s), consisting of two bright emission episodes separated by an inter-episode gap of $\sim600$~s with no emission detected above the GBM background. Using a novel HOP algorithm-based segmentation method, we identify optimal time intervals for time-resolved spectroscopy. A Comptonized power-law (CPL) model well describes the prompt emission ($\sim 10$~keV-–40 MeV) across all timescales.
\textit{Fermi}-GBM spectroscopy reveals a transition from a hard, slow-cooling synchrotron spectrum ($\alpha_{\rm I}=-0.70$) during episode--I to a softer, fast-cooling synchrotron spectrum ($\alpha_{\rm II}=-1.29$) during episode--II. This evolution closely resembles that of the magnetar-candidate GRB~091024A, highlighting a potential progenitor degeneracy based solely on spectral evolution.
However, the extreme isotropic-equivalent energy exceeds the maximum energy available from magnetar spin-down, strongly favouring a blue supergiant (BSG) collapsar origin. The prolonged activity and multi-episode structure are naturally explained by sustained accretion through an extended stellar envelope.
Our results demonstrate that ul-GRBs can display magnetar-like spectral evolution while being powered by BSG collapsars. GRB~220627A highlights the need to combine time-resolved spectroscopy with energetic constraints to distinguish among progenitor scenarios and to understand the physical origin of ul-GRBs.
\end{abstract}
%
}]




--------------------------------------------------
\section{Introduction}
\label{sec:Intro}

Among the many cosmic phenomena, Gamma-ray bursts (GRBs) release explosive, enormous energy in a brief amount of time. 
These events are marked by the rapid release of energy in the form of $\gamma$-ray photons and non-thermal radiation, over a short temporal interval.
These GRBs persist from milliseconds to hours, exhibiting isotropic luminosities of around $\sim 10^{51}-10^{54}~$erg/s (\cite{Zhang2018}).

GRBs are classified by $T_{90}$ duration (the time over which the burst emits $90\%$ of its total observed photons) into two primary categories: short GRBs (sGRBs) and long GRBs (lGRBs).
LGRBs occur when the massive star core collapses into a black hole or a rapidly rotating neutron star for a duration exceeding several seconds, even minutes \cite{Woosley1993, Paczynski1998}.
On the other hand, sGRBs result from compact-object binary mergers, which usually last up to $2~$s.
These bursts generate prompt emission from high-energy $\gamma-$ray photons originating in the central engine, produced by internal shocks. 
Following this, the ejected material interacts with its surroundings, producing an afterglow that persists from seconds to days across the electromagnetic spectrum (\cite{Levan2013, Abbott2017}).

Ultra-long GRBs (ul-GRBs) constitute a rare subclass of extended emissions, lasting from thousands of seconds to hours; however, their exact origin remains debated (\cite{Ioka2016}).
The recommended progenitors of ul-GRBs include blue supergiant (BSG), a result of prolonged high-energy emission of an extended envelope of a massive star (GRB 101225A, GRB 130925A) (\cite{Gendre2013, Kashiyama2013}); rapidly rotating magnetars, where due to elevated magnetic fields, a highly collimated jet is produced up to their spin-down lifetime (GRB 060218, GRB 100316D, and GRB 091024A) (\cite{Greiner2015, Gompertz2017}); tidal disruption events where white dwarfs are stripped off in the proximity of a supermassive black hole, supported by observational evidence Swift J1644+57 (\cite{Krolik2011, Komossa2015}).

The Fermi Gamma-ray Burst Monitor (GBM) (\cite{Meegan2009, Connaughton2015}) detected GRB~220627A on June 27, 2022, at $T_{0_{(1)}} = 21-21-0.086$~UTC; additionally, it was observed by the Fermi Large Area Telescope (LAT) (\cite{Atwood2009}).
Surprisingly, after $900~$s from the first trigger, at $T_{0_{(2)}} = 21-36-56.39$~UTC, GBM received another signal and restored the connection with a similar location and localised another long burst of GRB~220627A.
The $T_{90}$ duration of the trigger 220627890 observed by Fermi-GBM in $\gamma-$rays was detected around $138~$s ($10-1000~$keV), while the subsequent trigger 220627901 lasted only $127~$s ($10-1000~$keV)\footnote{We have excluded the energy channels in the range $31-35~$keV (due to the iodine K-edge at $33.17~$keV).}.
\cite{Roberts2022} hypothesised that this burst was gravitationally lensed, based on its temporal and spectral characteristics.
However, he left open the possibility that this GRB was a rare ul-GRB, given its exceptionally long detection period.


Following the dual Fermi-GBM triggers of GRB~220627A, a series of studies have progressively characterised this event across multiple wavelengths, yet several open questions remain.
\citet{Huang2022} performed the first joint GBM+LAT broadband spectral analysis, finding CPL+PL for episode--I (GeV photons present) versus CPL-only for episode--II (no LAT detection), ruling out gravitational lensing at $5.1\sigma$ and establishing this as the first ul-GRB with confirmed GeV emission. 
Their time-integrated approach, however, left the intra-episode spectral evolution and any photospheric contribution uncharacterised.

\citet{Wet2023} identified the optical afterglow, secured $z=3.084$ via VLT/MUSE, and modelled X-ray/optical/radio data within a single-component forward-shock framework ($\theta_j\approx4.5^\circ$, $n_0\approx20$\,cm$^{-3}$, $p\approx2.0$), concluding that the afterglow data require no unusual progenitor. 
\citet{Leung2026} subsequently showed via 7 to 456\,d radio monitoring (ATCA/ MeerKAT/ VLA/ VLBA) that a single forward shock cannot reproduce the late-time radio flattening after $\sim20$\,d, requiring an additional wide, slow cocoon ($E_{\rm cocoon}\sim10^{52}$\,erg) with a stellar radius $R_\star\sim10^{13}$\,cm, two orders of magnitude above a Wolf--Rayet star, and a cocoon energy exceeding the magnetar spin-down budget, strongly favouring a BSG collapsar.
The apparent tension between \citeauthor{Wet2023} (``no unusual progenitor required'') and \citeauthor{Leung2026} (``BSG progenitor required'') is complementary rather than contradictory: the former probed the early X-ray/ optical-dominated forward shock, while the latter revealed a radio-bright cocoon component only visible at late times and short radio wavelengths.
Resolving the progenitor picture, therefore, calls for independent diagnostics from the prompt emission itself.

This paper addresses this gap and answers the open questions that motivated our study. 
\begin{itemize}
    \item (i)~\textit{What sustains the central engine for $\gtrsim10^3$\,s?} \\
    Wolf--Rayet accretion timescales fall short by orders of magnitude \citep{Woosley1993, Ioka2016}, whereas a BSG envelope or magnetar spin-down can sustain activity over the required duration \citep{Metzger2007, Gendre2013};
    \item (ii)~\textit{How does energy dissipation evolve across the multi-episode prompt phase?} \\ 
    The $\sim10^3$\,s duration of GRB~220627A provides, for the first time in the ul-GRB cases, a sufficient temporal baseline to decompose the prompt emission into multiple spectral epochs and track the cooling regime transition \citep{Preece1998, Uhm2014};
    \item (iii)~\textit{Can progenitor degeneracy be broken by time-resolved spectroscopy?} \\ 
    Both BSG and magnetar models can produce similar time-integrated spectra \citep{Virgili2013, Perna2018}, but their spectral evolution and photospheric contributions may differ between episodes.
\end{itemize}
We address these questions by applying a novel HOP/Bayesian-block algorithm to time-resolved spectroscopy of GRB~220627A, and by comparing against the magnetar candidate GRB~091024A and BSG prototype GRB~130925A, the only other confirmed ul-GRBs with two Fermi-GBM-detected emission episodes suitable for direct comparison.

However, GRB~220627A is a rare GRB, as its $T_{100}$ spans $\sim 1100$~s and allows the study of spectral evolution through time-resolved spectroscopy. The work presented in this research provides a detailed time-resolved spectroscopic analysis using a state-of-the-art statistical technique, the HOP algorithm and Bayesian blocks. 
We have compared the time-resolved spectroscopic analysis of established GRB~220627A with those of the ul-GRBs 091024A $\&$ 130925A using the same Fermi-GBM data.
We probe the evolving energy-dissipation mechanisms that characterise such extraordinary events by dissecting the prompt emission phase at high temporal and spectral resolution. 

GRB~091024A and GRB~130925A were selected as the primary comparison sources because they are the only other confirmed ul-GRBs ($T_{90}\gtrsim10^3$\,s) with two spectroscopically distinct emission episodes well-detected by Fermi-GBM, whose spectral detection did not lead to upper limits in the second episode of Fermi-GBM or Fermi-LAT, enabling a uniform time-resolved CPL analysis in all three cases.
Within the current sample of confirmed ul-GRBs, \citep[see e.g.][]{Levan2013, Ioka2016, Leung2026}, eight events (excluding GRB 220627A) as of the time of writing \citep{Leung2026}, GRB~091024A is the canonical magnetar candidate \citep{Virgili2013}, while GRB~130925A is the best-established BSG collapsar \citep{Gendre2013, Evans2014, Piro2014}, making them natural end-member templates for the progenitor comparison.
GRBs with only single emission episodes (e.g.\ \citep[GRB~111209A][]{Gendre2013, Stratta2013}, \citep[GRB~121027A][]{Peng2013}, \citep[GRB~141121A][]{Cucchiara2015}, and \citep[GRB~170714A][]{Hou2018}) or insufficient temporal resolution of Fermi-GBM data among both episodes or defined progenitor status to match/compare time-resolved spectroscopy and progenitor study (e.g.\ \citep[GRB~150201A][]{Kienlin2020}, \citep[GRB~160625B][]{Zhang2018a}) were excluded.

A crucial diagnostic in the temporal domain is spectral lag, which quantifies the delay in the arrival of softer photons relative to harder ones (\cite{Norris2000, Norris2002, Norris2005}).
This delay time is computed using the Cross-Correlation Function (CCF); this lag reflects the temporal displacement between low- and high-energy light curves and encodes valuable information about the jet dynamics and the emission-region geometry (\cite{Fenimore1995}).
A natural explanation for photons originating from both episodes is that they are either the result of a single massive source or gravitationally lensed.

We have discussed data reduction for Fermi-GBM and LAT in Section~\ref{sec:Data_Reduc}.
The spectral evolution examined in Section~\ref{sec:Flux Variability} reveals variations in spectral parameters, consistent with changing microphysical conditions within the jet.
Section~\ref{sec:Time-Resolved Spectroscopy} discusses spectral changes over time-resolved spectroscopy for both episodes of GRB~220627A and a comparison study with known magnetar- and BSG-progenitor GRBs.
Section~\ref{sec:Spectral Study (GBM and LAT)} identifies significant parameter shifts across both emission episodes by describing in detail the GBM and LAT spectral contributions for thermal and non-thermal emissions.
Section~\ref{sec:Discussion} interprets the results obtained from the Fermi-GBM and Fermi-LAT analyses of all three ul-GRBs.
Appendix~\ref{sec:AppendixA} discusses the gravitational lensing properties compared with the progenitors of ul-GRBs via CCF and power spectral density (PSD) analysis, and Appendix~\ref{sec:AppendixB} confirms thermal signatures with thermal and a combination of thermal-non-thermal models\footnote{In this research, we utilise a Hubble constant of $H_0=71~$km/s/Mpc and cosmological parameters of $\Omega_m=0.27$ and $\Omega_\Lambda=0.73$.}.
%
%
\section{Data Reduction} \label{sec:Data_Reduc}
\subsection{GBM Data Reduction} \label{sec:GBM Data_Reduction}
The Fermi-GBM detected and localised GRB~220627A and transmitted the trigger information to further detectors and telescopes.
Fermi-LAT detected a GRB initially positioned at a boresight of $27^{\circ}$ when Fermi-GBM triggered (\cite{Lalla2022}).
Unexpectedly, nearly 600~s later, GBM re-established contact at a similar location and identified another prompt emission (\cite{Roberts2022}).
Despite favourable observational conditions and the absence of South Atlantic Anomaly (SAA) interference, the LAT instrument did not detect high-energy photons during the second emission episode.

We acquired GBM data from the \textit{HEASARC Archive}.\footnote{\url{https://fermi.gsfc.nasa.gov/ssc/data/access/gbm/}}
For our analysis of both episodes, temporally pre-binned Continuous Spectroscopy (CSPEC) and continuous Time-Tagged Event (TTE) datasets, each including $128~$energy channels, were selected, using a minimum bin size of $1024~$ms.
The GBM analysis utilised sodium iodide (NaI) and bismuth germanate (BGO) detectors with the lowest angle relative to the GRB's line of sight.
The episode--I was closely monitored by the closest viewing-angle detectors: n0, n1, n3 NaI, and b0 BGO, relative to the source direction, during its luminous phase from -6~s to 181.6~s relative to the trigger time $T_{0_{(1)}} (8~\mathrm{keV} - 40~\mathrm{MeV})$.
The episode--II burst duration was recorded by the n3, n6, n7 NaI and b1 BGO detectors, with the burst occurring between $-46~$s and $140~$s relative to the second trigger time $T_{0_{(2)}}$.

\begin{figure*}[hb!]
    \centering
    \includegraphics[width=0.7\textwidth]{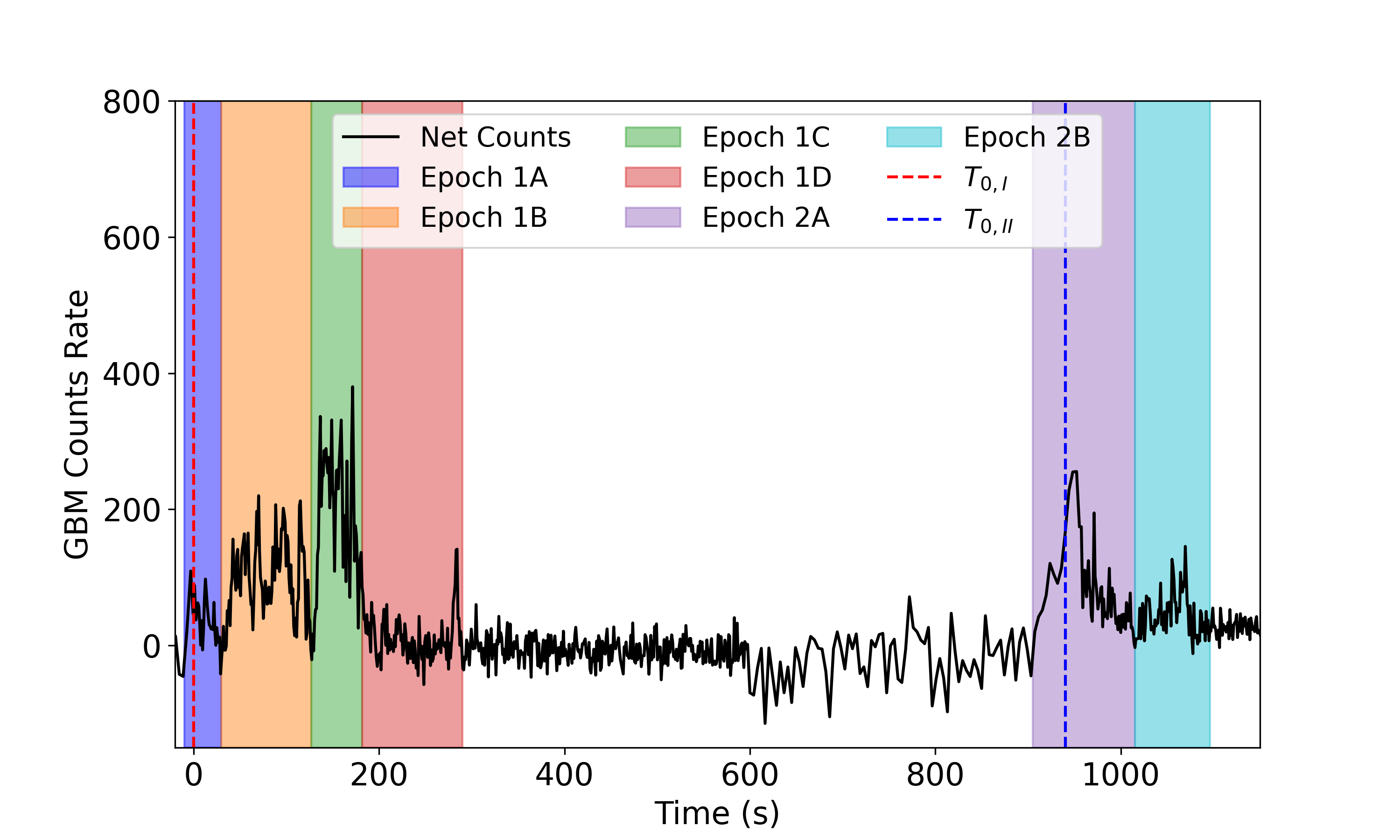} \\
    \caption{Total GBM light curve for the GRB~220627A showcasing all six epochs.}
    \label{fig:Lightcurve_TRS}
\end{figure*}

We performed background subtraction using polynomial fits of varying orders (first, second, and higher), as is standard in Fermi GBM analysis.
For this burst, a first-order polynomial yielded the lowest residuals and most reliable background determination for both time-resolved and time-integrated spectroscopy.
Additionally, we implemented a direction-dependent background fitting method (\cite{Szecsi2013}), which accounts for potential contamination from the Sun, Earth’s $\gamma$-ray emission, or other celestial sources along the line of sight of the Fermi-GBM detectors.
The light curve shown in Figure~\ref{fig:Lightcurve_TRS} resembles the structure of a bat, which is why we coined this source as `BAT GRB'.

We have applied the Eisenstein-Hut HOP blocks algorithm (\cite{Eisenstein1998, Wagner2021}) to the light curves of both episodes for all the ul-GRBs.
The algorithm identifies HOP blocks for all high-energy photons in the respective episode as discussed in Section~\ref{sec:Time-Resolved Spectroscopy}.
Upon applying all spectral functions, the CPL provided the ideal fit, which we will examine in depth in Section \ref{sec:Time-Resolved Spectroscopy}.
To ensure accurate error estimates of the spectral fit, we have used both asymmetric error propagation and Monte Carlo simulations.
The parameter evaluation to measure the statistical goodness-of-fit is estimated based on the ratio of the Profile-Gaussian likelihood statistic and degrees of freedom (PGStat/DOF) (\cite{Arnaud1996}).

To better understand ul-GRBs in the prompt emission domain, we have compared time-resolved spectroscopy of two known ul-GRB sources, GRB~091024A and GRB~130925A, using Fermi $\gamma-$ray data.
The Fermi-GBM detected and localised GRB~091024A at RA, DEC = $339.25^\circ,~56.89^\circ$ for the Fermi GBM triggers 091024372 and 091024380, respectively, observed using n7, n8, nb NaI, and b1 BGO detectors up to 100~s.
After approximately 631~s, another burst was detected for more than 400 s by n0, n6, n9, and b1 GBM detectors, as shown in the light curve~\ref{fig:Lightcurve_GRB}.
Fermi-GBM detected another ul-GRB source: GRB 130925A at RA, DEC $= 41.18^\circ,\ -26.13^\circ$ for the Fermi GBM triggers 130925164, 130925173, respectively, observed using n0, n1, n2 NaI, and b0 BGO detectors up to the initial 10~s.
Then, approximately $783\,\rm{s}$ later, the n6, n7, n9, b1 GBM detectors detected another burst for more than 250~s, as shown in the light curve~\ref{fig:Lightcurve_GRB}(a), (b), (c).
%
\subsection{LAT Data Reduction} \label{sec:LAT Data_Reduction}

The Fermi-LAT discovered GRB~220627A at RA, DEC$~=201.2^{\circ},~-32.5^{\circ}$ at a redshift of 3.084 and transmitted the trigger data to further telescopes. 
Data have been collected within a $10^{\circ}$ FOV of the Fermi-LAT telescope, centred on the GRB location, within a ($14^{\circ}\!\times\!14^{\circ}$) region of interest. 
We have performed an unbinned maximum-likelihood analysis utilising \texttt{LAT TRANSIENT} events (100~MeV to 300~GeV) using the \texttt{GTBURST}\footnote{\url{https://fermi.gsfc.nasa.gov/ssc/data/analysis/scitools/gtburst.html}} GUI software from the Fermi Science Tool, with a maximum zenith angle of $100^{\circ}$ to mitigate contamination from the $\gamma$-ray Earth limb. 
We have investigated up to 1100~s, but no high-energy photons have been observed after 700~s.
Unlike the time-integrated analysis of \cite{Huang2022}, which required a PL spectrum extension only for episode--I to fit LAT+GBM, our finer time-resolved GBM spectroscopy, detailed in section~\ref{sec:Time-Resolved Spectroscopy} using HOP blocks (Table~\ref{tab:GRB_TRS}), isolates CPL evolution without such components, highlighting intra-episode hardening/softening not resolved previously.

The analysis of Fermi-LAT data requires precise modelling of background contributions and instrument behaviour.
The galactic diffuse emission arising from cosmic ray interactions with interstellar gas and radiation is modelled using the file \texttt{gll\_iem\_v07.fits} \cite{LAT_Bkgd_Models}, which provides a spatial and spectral template for the Milky Way’s $\gamma$-ray foreground.
The isotropic background, which includes misclassified cosmic rays and unresolved extragalactic sources, is modelled using \texttt{iso\_P8R3\_TRANSIENT\_020\_V3\_v1.txt}, tuned to match the transient-class data used in this study.
To correctly interpret the LAT data, the Instrument Response Function (IRF) \texttt{P8R3\_TRANSIENT\_020\_V3} is employed. 
Using gtfindsrc, the best-fit position of GRB~220627A from Fermi-LAT data in the interval 0-700~s, assuming a PL spectrum, is RA = $201.181^\circ$ and DEC $= -32.4579^\circ$. 

\begin{table*}[hb!]
\centering
\renewcommand{\arraystretch}{1.2} 
\begin{threeparttable}
\begin{adjustbox}{width=\textwidth, center}
\begin{tabular}{ccccccc}
    \hline
    \textbf{Parameters} & \textbf{Epoch 1A} & \textbf{Epoch 1B} & \textbf{Epoch 1C} & \textbf{Epoch 1D} & \textbf{Epoch 2A} & \textbf{Epoch 2B}  \\ \hline
    \textbf{Episode} & I & I & I & I & II & II \\ 
    \textbf{Time Range (s)} ($\bm{T_{0_{(1)}}+}$) & -10 - 30 & 30 - 127 & 127 - 182 & 182 - 278 & 910 - 1015 & 1015 - 1096   \\
    \textbf{Peak Flux} \textbf{(ph/cm$^2$/s)} & $153.9 \pm 24.6$ & $367.6 \pm 26.6$ & $590.2 \pm 28.6$ & $162.6 \pm 24.6$ & $276.4 \pm 28.0$ & $274.0 \pm 28.0$  \\
    $\bm{\bar{x}}$ \textbf{(ph/cm$^2$/s)} & $57.3 \pm 6.3$ & $180.8 \pm 9.0$ & $328.6 \pm 18.3$ & $18.5 \pm 3.4$ & $107.6 \pm 5.6$ & $96.2 \pm 7.0$  \\
    $\bm{F_{var}}$ & $0.54 \pm 0.07$ & $0.46 \pm 0.03$ & $0.39 \pm 0.04$ & $-$\tnote{a} 
    & $0.47 \pm 0.04$ & $0.585 \pm 0.05$  \\
    \textbf{Remarks} & Flaring & Flaring & Flaring & Quiescent & Flaring & Flaring  \\
    \textbf{MVT (s)} & $3.42 \pm 0.79$ & $7.51 \pm 0.42$ & $3.61 \pm 0.40$ & $2.01 \pm 0.90$ & $6.78 \pm 0.94$ & $4.11 \pm 1.15$  \\ 
    \textbf{r} $\bm{(\times {10^{8}})~}$\textbf{cm} & $6.97 \pm 1.76$ & $15.33 \pm 1.76$ & $7.37 \pm 1.10$ & $4.10 \pm 1.89$ & $13.85 \pm 2.37$ & $8.40 \pm 2.49$  \\
    \textbf{R} $\bm{(\times {10^{15}})~}$\textbf{cm} & $4.52 \pm 1.38$ & $9.93 \pm 2.06$ & $4.77 \pm 1.09$ & $2.65 \pm 1.30$ & $8.97 \pm 2.18$ & $5.44 \pm 1.87$ \\ 
    \hline
    
\end{tabular}
\end{adjustbox}
\begin{tablenotes}
\footnotesize
\item[a] $F_{\rm var}$ is not reported: $\sigma^2_{\rm XS}=570\pm78\,(\mathrm{ph\,cm^{-2}\,s^{-1}})^2$ is too large for a reliable value (see §\ref{sec:gamma-ray_Variability}).
\end{tablenotes}
\caption{Temporal and flux variability parameters for selected epochs of both emission episodes of GRB~220627A using GBM data from average NaI detectors. The table summarises the mean photon flux ($\bar{x}$), fractional variability ($F_{var}$), MVT, estimated size of the $\gamma$-ray emitting blob (r) and the distance of the $\gamma$-ray emission blob to the central engine (R) in the observer frame. Epochs 1A to 1D belong to episode--I, and epochs 2A and 2B belong to episode--II. Epoch 1D is classified as the post-flare/quiescent interval.}
\label{tab:flux_variability}
\end{threeparttable}
\end{table*}

\section{Results} \label{sec:Results}

The GRB~220627A presents a rare opportunity to investigate either an ul-GRB or a potential gravitationally lensed event.
To characterise its emission properties, we performed a flux-variability analysis and time-resolved spectroscopic study to probe the source's evolution across different phases.
We utilised the spectral analysis tools in Gamma-ray Data Tools (GDT) (\cite{GbmDataTools}) to examine the full $T_{90}$ duration, leveraging TTE data from both episodes.

\subsection{Flux Variability} \label{sec:Flux Variability}
\subsubsection{Gamma-ray Variability} 
\label{sec:gamma-ray_Variability} 

We have analysed the flux variability for all six prominent epochs in the entire light curve: \textit{1A, 1B, 1C, 1D, 2A, and 2B}, as summarised in Table~\ref{tab:flux_variability} for the epochs shown in Figure~\ref{fig:Lightcurve_TRS}.
Peak fluxes and their statistical uncertainties are reported for each epoch at the $1\sigma$ level.
To quantify the intrinsic variability of the burst, we computed the fractional variability amplitude ($F_{var}$):
\begin{equation}
    F_{\rm var} = \sqrt{\frac{S^2 - \overline{\sigma^2_{\rm err}}}{\overline{x^2}}}
    \label{eq:Fvar}
\end{equation}
based on equations 6, 9, B2, and 10 of \cite{Vaughan2003}, where $S^2$ is the sample variance, $\overline{\sigma^2_{\rm err}}$ is the mean square error, and $\bar{x}$ is the arithmetic mean of the light curve.
We have ensured that the observed variability is statistically significant, as stated in (\cite{Vaughan2003}). 
The relative amplitude of flux variability for the flaring regions 1A, 1B, 1C, 1D, 2A, and 2B is confined to a narrower band of $F_{var} = 0.4-0.6$.
Epoch~1D is excluded from this comparison as the flux is too low for a reliable $F_{var}$ measurement.


Epoch~1D ($182$--$278$\,s) spans the post-flare quiescent tail of episode--I, coinciding with the onset of the $\sim600$\,s inter-episode gap. 
It is retained in the time-resolved spectroscopy (Table~\ref{tab:GRB_TRS}) to probe spectral evolution during the quiescent phase relative to the preceding flaring epochs, and in the Fermi-LAT analysis to assess whether any high-energy emission persists into this interval. 
The time-integrated spectral fit in Table~\ref{tab:GBM_full_spectral} uses the $T_{90}$ interval ($-6$\.s to $+181.6$\,s) only and excludes this quiescent tail.

For the flaring epochs~1A--1C, the net counts per bin exceed the $1 \sigma$ count-rate uncertainty by factors of $3-5$, confirming that the measured $F_{\rm var}=0.4-0.6$ reflects genuine source variability. 
In epoch~1D, by contrast, the net counts in individual bins are consistent with the background uncertainty at less than $2 \sigma$, rendering the $F_{\rm var}$ statistic unreliable regardless of whether the terminal spike is included or excluded,
in both cases, the value unphysically exceeds unity. 
Following \citet{Vaughan2003} (their Appendix~B), when $\overline{\sigma^2_{\rm err}}/\bar{x}^2 \gg1$ the fractional rms becomes untrustworthy; we therefore do not report $F_{\rm var}$ for epoch~1D and instead note that the excess variance $\sigma^2_{\rm XS} = S^2 - \overline{\sigma^2_{\rm err}} = 570\pm78\, (\rm ph\,cm^{-2}\,s^{-1})^2$ is reported for completeness, at this signal-to-noise, we do not interpret it as intrinsic variability.


To probe whether a quiescent flux bridges the two episodes, we examined the NaI n3 light curve, the only detector with a favourable viewing geometry for both triggers (mean angle $13.9^\circ$, minimum $2.5^\circ$; §\ref{sec:GBM Data_Reduction}), with the direction-dependent background fitting method of \citep{Szecsi2013}. 
In the inter-episode gap ($\sim290-910$\,s relative to $T_{0_1}$), the background-subtracted n3 count rate is consistent with zero 
(mean residual $< 2$\,counts per 4.1\,s bin, against the episode--I peak of $\sim350$\,counts/bin), with no statistically significant quiescent emission bridging the episodes.
The LAT energy flux in epochs~1C and~1D (detailed in \S\ref{sec:Fermi-LAT Lightcurve and Spectra}) declines with a temporal index $\alpha_t = 2.0 \pm 1.1$ (Figure~\ref{fig:LAT_220627A}b), steeper than the canonical afterglow decay ($\alpha_t \sim  1.2$; \citealt{Zhang2006, Nousek2006}), though with two points, this is only indicative. 
The first confirmed afterglow detection (optical/X-ray/radio) occurs only after Episode~II concludes \citep{Wet2023, Leung2026}.
Both the GBM non-detection and the steep LAT decay are consistent with rapid central engine cessation during the inter-episode gap, supporting the BSG fallback accretion picture in which the accretion rate drops sharply as the jet encounters density discontinuities in the stratified stellar envelope \citep{Nakauchi2013, Perna2018}.

\subsubsection{Gamma-ray emitting region} 
\label{sec:gamma-ray_emitting_region}

The MVTs established a connection between the distance the jet travels radially before emitting $\gamma$-rays (R) (\cite{Dermer2004, Nousek2006, Zhang2006, Prince2019}) and the maximum size of the causally connected $\gamma$-ray emitting blob at radius R, (r) (\cite{Sari1998, Granot2008}), given by the relation 
\begin{equation}
\begin{split}
    R \approx 2 \Gamma^2 c \mathrm{MVT}/(1+z) \\
    r \leq c \mathrm{MVT} (1+z)/ \delta,
\end{split}
\label{eq:r-R}
\end{equation}
where $\Gamma \geq 300$ (adopted from (\cite{Huang2022}), typical for GRBs, $\delta \approx 2\Gamma$ (on-axis; \cite{Lithwick2001}).
We detect the narrowest pulse with a full width at half maximum (FWHM) that is above the background and satisfies SNR thresholds with the multiple excess peak search algorithm (\cite{Guidorzi2015}) and consider the minimum FWHM of statistically significant pulses ($\mathrm{SNR}~>~3\sigma$) to be the MVTs (\cite{MacLachlan2013, Greiner2014, Camisasca2023}). 
\begin{figure}[H]
    \centering
    \includegraphics[width=0.45\textwidth, height=0.48\textwidth]{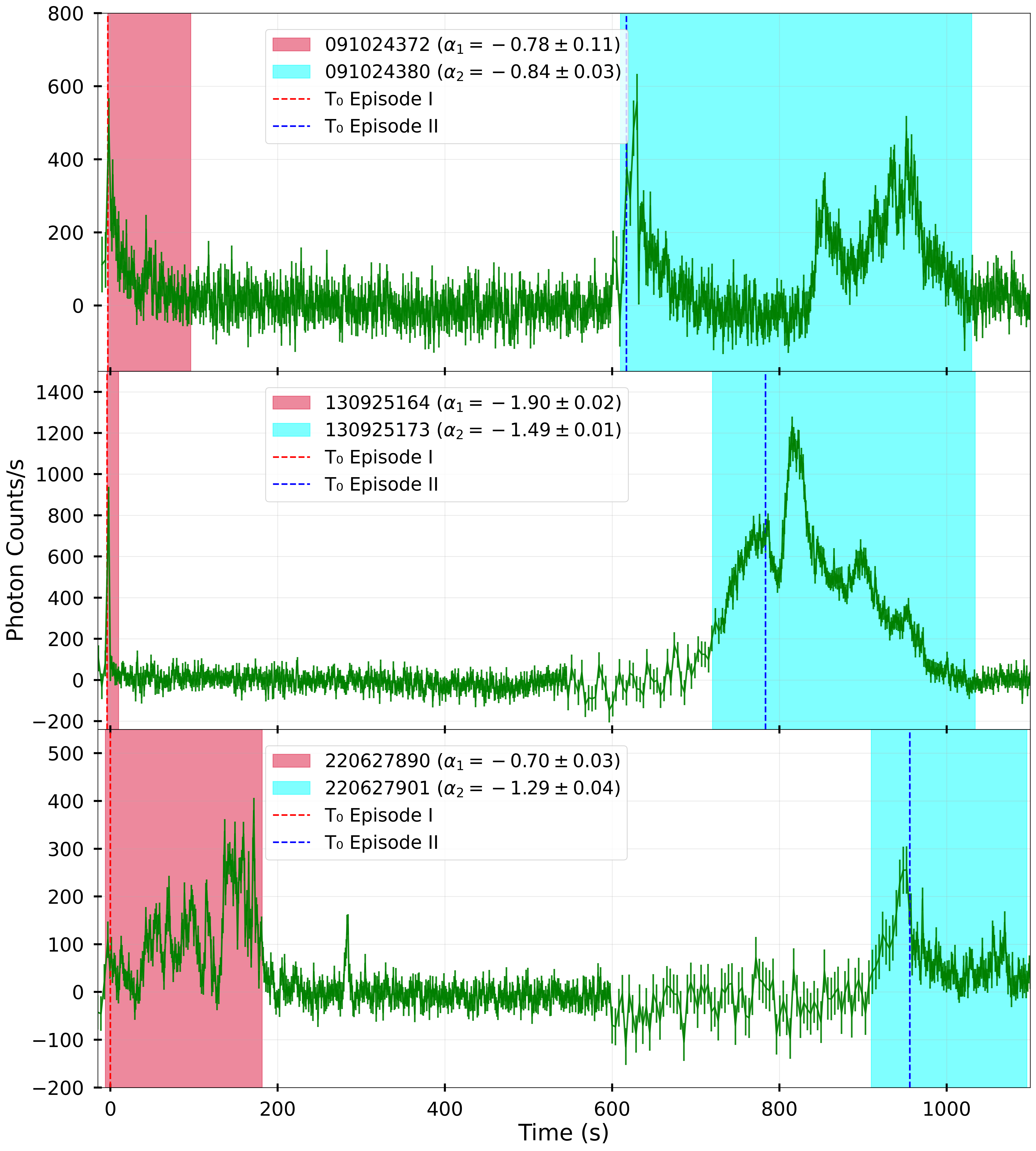} \\
    \small (a) Ul-GRB Light curve \\
    \vspace{10pt} 
    
    \includegraphics[width=0.45\textwidth, height=0.26\textwidth]{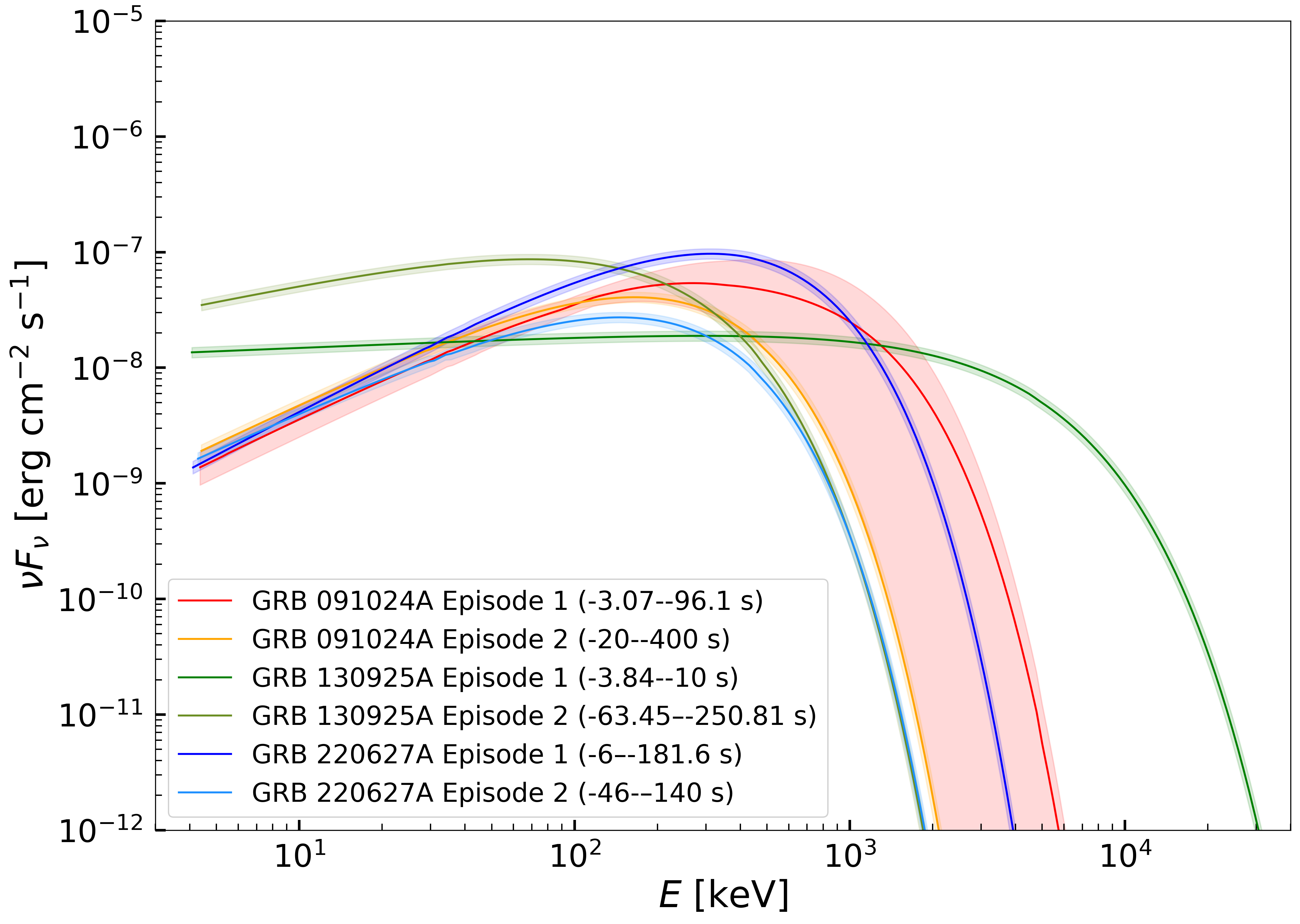} \\
    \small (b) Ul-GRB SED \\
    \vspace{10pt} 
    \caption{(Top plot) Combined light curve for GRB 091024A, 130925A, and 220627A. The background-subtracted GBM count rate, with tomato and cyan shaded regions marking T90 periods for episodes-I and II, respectively. Vertical dashed lines labelled $T_{0_{(1)}}$ and $T_{0_{(2)}}$ indicate the trigger times for each episode. (Bottom plot) SED plot for both episodes of all ul-GRBs for their T90 time duration: red for 091024372, orange for 091024380, green for 130925164, light green for 130925173, blue for 220627890, light blue for 220627901, respectively.}
    \label{fig:Lightcurve_GRB}
\end{figure}


\begin{table*}[t!]
\centering
\renewcommand{\arraystretch}{1.5} 
\begin{adjustbox}{width=0.9\textwidth,center}
\begin{tabular}{cccccccc}
    \hline

    \textbf{Source Range} & $\bm{E_{(p, I)}}$ & $\bm{\alpha_I}$ & \textbf{PGStat\_I/} & \textbf{Source Range} & $\bm{E_{(p, II)}}$ & $\bm{\alpha_{II}}$ & \textbf{PGStat\_II/} \\ 
    (s) & \textbf{(keV)} &  & \textbf{DOF\_I} & (s) & \textbf{(keV)} &  & \textbf{DOF\_II} \\
    \hline
    
    \multicolumn{4}{c}{\textbf{GRB~220627A Episode 1}} & \multicolumn{4}{c}{\textbf{GRB~220627A Episode 2}} \\ \hline

    (-10, 30) & $206.03^{+19.97}_{-20.86}$ & $-0.44^{+0.139}_{-0.163}$ & $135.97/116$ & (-46, 59) & $374.63^{+60.09}_{-78.02}$ & $-1.32^{+0.050}_{-0.055}$ & $173.22/116$  \\ 
    (30, 127) & $297.30^{+9.46}_{-9.69}$ & $-0.70^{+0.037}_{-0.039}$ & $149.04/116$ & (59, 140) & $136.01^{+13.53}_{-15.01}$ & $-1.18^{+0.046}_{-0.049}$ & $128.13/116$  \\ 
    (127, 182) & $365.87^{+10.60}_{-10.85}$ & $-0.72^{+0.033}_{-0.034}$ & $181.44/116$ &  &  &  &    \\ 
    (182, 278) & $381.90^{+147.69}_{-308.91}$ & $-1.07^{+0.025}_{-0.038}$ & $138.57/116$ &  &  &  &    \\ 
    \hline
    

    \multicolumn{4}{c}{\textbf{GRB~091024A Episode 1}} & \multicolumn{4}{c}{\textbf{GRB~091024A Episode 2}} \\ \hline
    (-10, 19.7) & $434.54^{+43.55}_{-47.96}$ & $-0.74^{+0.109}_{-0.126}$ & $134.86/119$ & (-16.385, -1.025) & $202.88^{+57.51}_{-87.41}$ & $-1.29^{+0.10}_{-0.12}$ & $480.60/503$   \\ 
    (40.2, 46.4) & $238.38^{+45.08}_{-53.23}$ & $-0.79^{+0.166}_{-0.209}$ & $81.47/119$ & (0.255, 23.55) & $297.86^{+51.71}_{-67.71}$ & $-1.29^{+0.06}_{-0.06}$ & $693.65/503$  \\ 
     &  &  &  & (205.83, 261.125) & $110.83^{+8.03}_{-8.64}$ & $-1.26^{+0.030}_{-0.031}$ & $535.39/505$  \\ 
     &  &  &  & (261.125, 350) & $245.15^{+14.63}_{-15.99}$ & $-1.32^{+0.022}_{-0.022}$ & $1635.62/505$  \\ 
     \hline

    \multicolumn{4}{c}{\textbf{GRB~130925A Episode 1}} & \multicolumn{4}{c}{\textbf{GRB~130925A Episode 2}} \\ \hline
    (-11, 0) & $9.04^{+3.46}_{-6.62}$ & $-1.90^{+0.0001}_{-0.044}$ & $261.09/503$ & (-5, 65) & $198.09^{+7.41}_{-7.79}$ & $-1.35^{+0.011}_{-0.011}$ & $1070.89/501$ \\ 
    (0, 10) & $20.59^{+4.54}_{-4.56}$ & $-0.68^{+0.104}_{-0.150}$ & $285.89/503$ & (70, 100) & $84.63^{+3.14}_{-3.15}$ & $-1.55^{+0.008}_{-0.008}$ & $1305.91/501$ \\ 
     &  &  &  & (150, 200) & $37.12^{+1.11}_{-1.14}$ & $-1.38^{+0.009}_{-0.009}$ & $1043.73/501$ \\ 
     &  &  &  & (200, 280) & $36.07^{+1.74}_{-1.82}$ & $-1.34^{+0.014}_{-0.014}$ & $458.47/501$ \\ \hline
    
\end{tabular}
\end{adjustbox}
\caption{Time-resolved spectral parameters for episodes 1 and 2 of GRB~220627A, GRB~091024A, and GRB~130925A using GBM data, fitted with a CPL model. The table lists the source intervals, the spectral peak energy ($E_p$) in keV, the isotropic-equivalent luminosity in erg/s, the low-energy spectral index ($\alpha$), and the PGStat/DOF values for episodes 1 and 2.}
\label{tab:GRB_TRS}
\end{table*}

\begin{table*}[t!]
\centering 
\renewcommand{\arraystretch}{1.2} 
\begin{adjustbox}{width=\textwidth,center} 
\begin{tabular}{>{\centering\arraybackslash}m{2.5cm}>{\centering\arraybackslash}m{1.5cm}>{\centering\arraybackslash}m{2cm}>{\centering\arraybackslash}m{2cm}>{\centering\arraybackslash}m{2.4cm}>{\centering\arraybackslash}m{3.5cm}>{\centering\arraybackslash}m{4.5cm}}
\hline
\textbf{Property} & $\bm{E_p}$ \textbf{(keV)} & $\bm{E_{\textbf{iso}}}$ \textbf{(erg)} & $\bm{T_{90}}$ \textbf{(s)} & \textbf{LCM} & \textbf{Spectral Signatures} & \textbf{References}   \\ \hline
\textbf{BSG Collapsar} & $> 100$ & $\sim10^{54}$ & $\sim 10^3 - 10^4$ & Multi-peaked flares & Thermal Emission, Structured Variability & \cite{Gendre2013, Evans2014, Piro2014, Perna2018}   \\ \hline
\textbf{Newborn Magnetar} & Variable & $\sim 10^{53} - 10^{54}$ & $\sim 10^4 - 10^6$ & Smooth Spin-down & X-ray plateaus & \cite{Metzger2007, Virgili2013, Lue2017, Fu2024, Ror2024}    \\ \hline
\textbf{WD-TDE} & $< 100$ & $\sim10^{53}$ & $\sim 10^4 - 10^5$ & Flares + Days Extended & Ionized ejecta & \cite{Krolik2011, Gendre2013, Levan2013, Komossa2015, Ioka2016} \\ \hline
\textbf{GRB 220627A} & $\sim 249$ & $\sim 1.6 \times 10^{54}$ & $1.09 \times 10^3$ & Multi-peaked & Thermal emission, Structured Variability & \textbf{This paper}   \\ \hline
\textbf{GRB 091024A} & $\sim 300$ & $\sim 1.9 \times 10^{53}$ & $1.01 \times 10^3$ & Inter-pulse emission & X-ray, optical plateaus & \cite{Virgili2013}    \\ \hline
\textbf{GRB 130925A} & $\sim 297$ & $\sim 1.15 \times10^{52}$ & $1.05 \times 10^3$ & Multi-peaked, Dust-echo  & Thermal X-ray spectrum, afterglow & \cite{Evans2014, Piro2014}    \\ \hline
\end{tabular}
\end{adjustbox}
\caption{Prompt emission properties of GRB 220627A compared to ul-GRB progenitor models, where $E_{iso}$ is the total isotropic energy (in erg), $E_p$ is the SED peak energy (in erg), $T_{90}$ denotes the duration required to achieve 90$\%$ fluence (in s), LCM is the light curve morphology, and spectral signatures of the progenitor.} 
\label{tab:multi-model}
\end{table*}

\subsection{Time-resolved Spectroscopy} \label{sec:Time-Resolved Spectroscopy}
In this study, the spectral evolution of GRB~220627A is examined by dissecting the light curve into smaller time bins using Bayesian blocks. 
We have used Fermi-GDT (\cite{GbmDataTools}) to obtain the best spectral model: CPL, whose details are shared in Section~\ref{sec:Data_Reduc}

\subsubsection{GRB~220627A: Ultra-long or Gravitationally lensed:} \label{sec:Ultra-long or gravitationally lensed}

The study of spectral evolution in short/long GRBs is difficult, but ul-GRBs provide enough temporal room to do it. 
To understand the spectral evolution, we applied the HOP algorithm, which created a Bayesian block (\cite{Eisenstein1998, Wagner2021}) in both episodes, but to maintain coherence, we kept the time bins from flare start to fall.

Using this approach, we have divided the first 300~s of episode--I into four epochs and the first 150~s of episode--II into two epochs, as shown in Figure~\ref{fig:Lightcurve_TRS}.
To enable intra-episode spectral tracking, Bayesian blocks are generated for all six high-flux states, 
spanning a 1-s bin (Table~\ref{tab:multi-model}), a time window not accessible in time-integrated studies.
We have fitted the CPL function to episode--I and episode--II to obtain spectral features, including the Spectral Energy Distribution (SED) peak, spectral index, photon flux, and the PGStat/DOF ratio.

The episode--I spectrum is more complex, hardening as it reaches the higher-flux states, while episode--II tends to show spectrum softening.
The spectral index for episode--I is the hardest-observed, at $-0.44$ in epoch 1A; contrastingly, episode--II shows a softer spectrum with a spectral index of $-1.3$ in epoch 2A, suggesting no spectral resemblance between the two episodes (Table~\ref{tab:GRB_TRS}).

\subsubsection{Comparison of spectrum with known ul-GRBs}
\label{sec:Multi-Model Progenitor Analysis (ul-GRBs)}

We compare the episode-resolved GBM spectra of GRB~220627A, 091024A, and 130925A against established ul-GRB progenitors: BSG collapsars, rapidly rotating newborn magnetars, and white-dwarf tidal disruption events (WD-TDEs). 
Selected time intervals from section~\ref{sec:Flux Variability} in both episodes yield CPL parameters:

\begin{table*}[ht!]
\renewcommand{\arraystretch}{1.5} 
\centering
\begin{adjustbox}{width=\textwidth, center}
\begin{tabular}{ccccccccc}
\hline
    \textbf{Model} & $\mathbf{E_p}$ & \textbf{kT} & $\boldsymbol{\alpha}$ & \textbf{PGStat/} & $\mathbf{E_{\textbf{Flux}}}$ ($\boldsymbol{\times 10^{-7}}$) & $\mathbf{Ph_{\textbf{Flux}}}$ & \textbf{Iso. Luminosity} & \textbf{Iso. Energy} \\
     & \textbf{(keV)} & \textbf{(keV)} &  & \textbf{DOF} & (\textbf{erg/s/cm$^2$}) & (\textbf{ph/s/cm$^2$}) & ($\boldsymbol{\times 10^{52}}$ \textbf{erg/s)} & ($\boldsymbol{\times 10^{54}}$ \textbf{erg)} \\    \hline
    \multicolumn{9}{c}{\textbf{GBM Episode - I ($\mathbf{T_{0_{(1)}} -6 \to T_{0_{(1)}} + 181.6}$) }}\tabularnewline
    \hline
    CPL & $317.21^{+6.85}_{-6.96}$ & - & $-0.70 \pm {0.03}$ & $267.39/116$ & $3.38 \pm 0.1$ & $1.73 \pm 0.05$ & $2.90 \pm 0.24$ & $ 1.33 \pm 0.11$  \\ 
    BB & - & $29.99^{+0.01}_{-1.05}$ & - & 
    - & $1.72 \pm 0.13$ & $1.30 \pm 0.07$ & $1.48 \pm 0.16$ & $0.68 \pm 0.07$  \\ 
    CPL+BB & $408.44^{+13.60}_{-17.97}$ & $30.41^{+1.08}_{-1.18}$ & $-0.82 \pm 0.04$ & $247.86/ 114$ & $3.52 \pm 0.15$ & $1.73 \pm 0.07$ & $3.02 \pm 0.27$ & $1.39 \pm 0.12$  \\ 
    \hline

    \multicolumn{9}{c}{\textbf{GBM Episode - II ($\mathbf{T_{0_{(2)}} -46 \to T_{0_{(2)}} + 140}$) }}\tabularnewline
    \hline
    CPL & $231.35^{+22.05}_{-24.98}$ & - & $-1.29 \pm {0.04}$ & $226.02/116$ & $1.42 \pm 0.12$ & $1.23 \pm 0.08$ & $1.22 \pm 0.14$ & $ 0.56 \pm 0.06$ \\ 
    BB & - & $29.99^{+0.73}_{-0.12}$ & - & 
    - & $0.84 \pm 0.06$ & $0.63 \pm 0.04$ & $0.72 \pm 0.07$ & $0.33 \pm 0.03$  \\ 
    CPL+BB & $432.31^{+137.26}_{-426.37}$ & $31.34^{+1.67}_{-1.50}$ & $-1.60 \pm 0.05$ & $217.68/ 114$ & $1.43 \pm 0.01$ & $1.25 \pm 0.01$ & $1.23 \pm 0.09$ & $0.56 \pm 0.04$  \\ 
    \hline

    
    
    
    \hline
    \hline
    \multicolumn{9}{c}{\textbf{LAT Emission ($\mathbf{T_{0_{(1)}} + 0 \to T_{0_{(1)}} + 700}$) }}\tabularnewline
    \hline
    \textbf{Model} &  &  & $\boldsymbol{\lambda}$ & \textbf{TS} & $\mathbf{E_{\textbf{Flux}}}$ ($\boldsymbol{\times 10^{-8}}$) & $\mathbf{Ph_{\textbf{Flux}}}$  ($\boldsymbol{\times 10^{-5}}$) & \textbf{Iso. Luminosity} & \textbf{Iso. Energy} \\
    &  &  &  &  & (\textbf{erg/s/cm$^2$}) & (\textbf{ph/s/cm$^2$}) & ($\boldsymbol{\times 10^{51}}$ \textbf{erg/s)} & ($\boldsymbol{\times 10^{53}}$ \textbf{erg)} \\    \hline
    PL & - & - & $-2.18 \pm 0.15$ & 192 & $1.37 \pm 0.45$ & $1.69 \pm 0.26$ & $1.18 \pm 0.40$ & $2.02 \pm 0.68$ \\ \hline
\end{tabular}
\end{adjustbox}
\caption{GBM spectral parameters for both episodes for their respective time duration for GRB~220627A. The CPL, BB and CPL+BB models are shown with their parameters: SED Peak Energy ($E_p$) (in keV), Low-energy spectral index ($\alpha$), thermal temperature ($kT$) (in keV), PL spectral index ($\lambda$), PG Statistics (PGStat), Test Statistics (TS), degrees of freedom DOF, Energy Flux ($E_{Flux}$) (in erg/s/cm$^2$), Photon Flux ($Ph_{Flux}$) (in ph/s/cm$^2$), isotropic luminosity (in erg/s) and isotropic energy (in erg).}
\label{tab:GBM_full_spectral}
\end{table*}

SED peak energy ($E_p$), isotropic-equivalent luminosity ($L_{iso}$), and low-energy spectral index ($\alpha$) (Table~\ref{tab:GRB_TRS} and Figure~\ref{fig:Lightcurve_GRB}(b)). 
In GRB~091024A and 130925A, the initial onset is not sufficient to discuss the spectral evolution when compared with ul-GRBs. 
However, GRB~220627A offers us an exclusive opportunity to study the longer-term evolution of the spectral index inferred from the CPL model.

GRB~220627A and GRB~091024A both display harder episode--I to softer episode--II evolution (220627A: $\alpha_I \sim -0.44$ to $-1.07 \to \alpha_{II} \sim -1.32$ to $-1.18$ vs 091024A\footnote{GRB~091024A episode--II combines the literature Episodes-II and III (\cite{Virgili2013}). This maintains consistent episode comparison across all three ul-GRBs.}: $\alpha_I \sim -0.74$ to $-0.79 \to \alpha_{II} \sim -1.29$ to $-1.32$). Spectroscopic results reveal progenitor degeneracy among consecutive episodes of all three GRBs (Figure~\ref{fig:Lightcurve_GRB}(b)). The spectra of episode--I of GRB~220627A closely match Episode-I of magnetar candidate GRB~091024A, while the low-energy spectral index of episode--II resembles episode--II of GRB~091024A (magnetar); however, the curvature is more aligned with episode--II of GRB~130925A (BSG). This contradiction prevents either progenitor model from uniquely explaining GRB~220627A's spectral evolution.

\begin{table*}[ht!]
\centering
\small
\renewcommand{\arraystretch}{1.} 
\begin{adjustbox}{width=0.9\textwidth, center}
\begin{tabular}{lccc}
    \hline
    \textbf{GRB} & \textbf{kT (keV)} & \textbf{Properties} & \textbf{Reference} \\
    \hline
    GRB 060218 & $\sim~10-20$ & Jet breakout +  & \cite{Ghisellini2007} \\
               &             & nearby SN Ib & \\
    GRB 090902B & $\sim~15-30$ & Photospheric +  & \cite{Abdo2009} \\
                &             & internal shocks & \\
    GRB 110721A & $\sim~20-40$ & Hybrid thermal + & \cite{Axelsson2012} \\
                &             & non-thermal spectrum & \\
    GRB 130925A & $\sim~10-50$ & Ul-GRB (multi-episodes), & \cite{Piro2014} \\
                &             & jet-cocoon signatures &  \\
    GRB 160625B & $\sim~10-20$ & Synchrotron-like +  & \cite{Lue2017} \\
                &             & thermal in prompt emission &  \\
    GRB 220627A & $\sim~20-31$ & Ul-GRB, jet-cocoon + & This work \\
                &             & photospheric emission & \\
    \hline
\end{tabular}
\end{adjustbox}
\caption{Thermal components detected in multi-episode and ul-GRBs with their thermal temperature $kT$ (in keV), their respective properties, along with the known literature assistance.}
\label{tab:thermal_grbs}
\end{table*}

\section{Spectral Study of GRB~220627A (GBM and LAT)} \label{sec:Spectral Study (GBM and LAT)}
\subsection{GBM Time-Integrated Spectra} \label{sec:GBM Time-Integrated Spectra}

For the time-integrated prompt emission of GRB~220627A, we fitted the GBM spectra of Episodes I and II, as well as the total burst duration, with the CPL model (Table~\ref{tab:GBM_full_spectral}). 
Episode--I $(-6~\mathrm{s},~181.6~\mathrm{s})$ is characterised by a peak energy $E_p~=~317.21^{+6.85}_{-6.96}$~keV and low-energy spectral index $\alpha_{\mathrm{GBM,1}}~=~-0.70~\pm~0.03$, while episode--II $(-46~\mathrm{s},~140~\mathrm{s})$ shows a softer spectrum with $E_p~=~231.35^{+22.05}_{-24.98}$~keV and $\alpha_{\mathrm{GBM,2}}~=~-1.29~\pm~0.04$. 
The total GBM emission integrated over -6~s to 1096~s yields $E_p~=~248.90^{+8.28}_{-8.50}$~keV and $\alpha~=~-0.81~\pm~{0.04}$, with an isotropic-equivalent energy $E_{\mathrm{iso}}~=~1.56~(\pm~0.14)\times10^{54}$~erg at $z~=~3.084$, placing GRB~220627A among the most energetic ul-GRBs reported to date, with BSG candidates GRB~101225A, GRB~111209A/ SN~2011kl, and GRB~130925A (\cite{Gendre2013, Levan2013, Evans2014, Piro2014, Greiner2015}).

The isotropic-equivalent luminosity peaks at $L_{\mathrm{iso,~I}}~=~2.90~(\pm~0.24)\times 10^{52}$~erg/s for episode--I and $L_{\mathrm{iso,~II}}~=~1.22~(\pm~0.14)\times10^{52}$~erg/s for episode--II (Table~\ref{tab:GBM_full_spectral}), confirming that episode--I is both spectrally harder and intrinsically more luminous. 
These values are consistent with the time-resolved HOP analysis in Section~\ref{sec:Multi-Model Progenitor Analysis (ul-GRBs)}.

The time-integrated properties favour BSG, GRB~220627A's $E_{\rm iso} = 1.6\times10^{54}$~erg exceeding 091024A ($1.9\times10^{53},\,\sim8$ times lower) and 130925A ($1.15\times10^{52},\,\sim13$ times lower) (Table~\ref{tab:multi-model}), 
with 600~s quiescence, thermals ($kT=22-31$~keV (sec.~\ref{sec:Thermal Component (CPL+BB)}), radio cocoon $R_{ph} \approx 1.3\times10^{13} (M/10M_\odot)^{-1/2}$~cm (\cite{Leung2026}) matching cocoon $R_{ph,c}$ (\ref{sec:AppendixB})), beyond magnetar spin-down limits ($\sim10^{53}$~erg).

\subsection{Thermal Component (CPL+BB)} \label{sec:Thermal Component (CPL+BB)}

The lower-energy systematic flattening/soft excess below $E_p$ in several time bins suggests a thermal component not fully captured by the CPL model alone.
To investigate this, we performed a separate analysis of flaring time bins by introducing an additional black-body (BB) component to test for photospheric emission. 
\ref{sec:AppendixB} illustrates a detailed BB/CPL+BB analysis confirming thermal detection ($kT\sim22-31$~keV) with episode--resolved evolution (Figure~\ref{fig:GBM_thermal}).
The CPL+BB model with 31-35~keV exclusion (Table~\ref{tab:GRB_TRS}) shows improved fits in the time bin of episode--I, where the best-fit kT values lie above or below the K-edge-affected range. 


For Table~\ref{tab:thermal_grbs}, the GRBs were selected on the criterion of a confirmed spectral thermal component ($kT\sim10-50$ keV) in the prompt or early afterglow emission, drawing from published analyses of Fermi and Swift data.
This yields a heterogeneous but physically motivated sample spanning classical long GRBs with photospheric emission (GRB~060218, GRB~090902B, GRB~110721A, GRB~160625B) and ul-GRBs with jet-cocoon thermal signatures (GRB~130925A, GRB~220627A), allowing us to place GRB~220627A's $kT\sim22$--$31$\,keV squarely within the empirical range expected from BSG cocoon models \citep[kT$\sim$10--100\,keV;][]{Lazzati2017}.

\subsection{Fermi-LAT Light curve and Spectra of GRB~220627A} \label{sec:Fermi-LAT Lightcurve and Spectra}

\begin{figure}[H]
    \centering
    \includegraphics[width=0.45\textwidth]{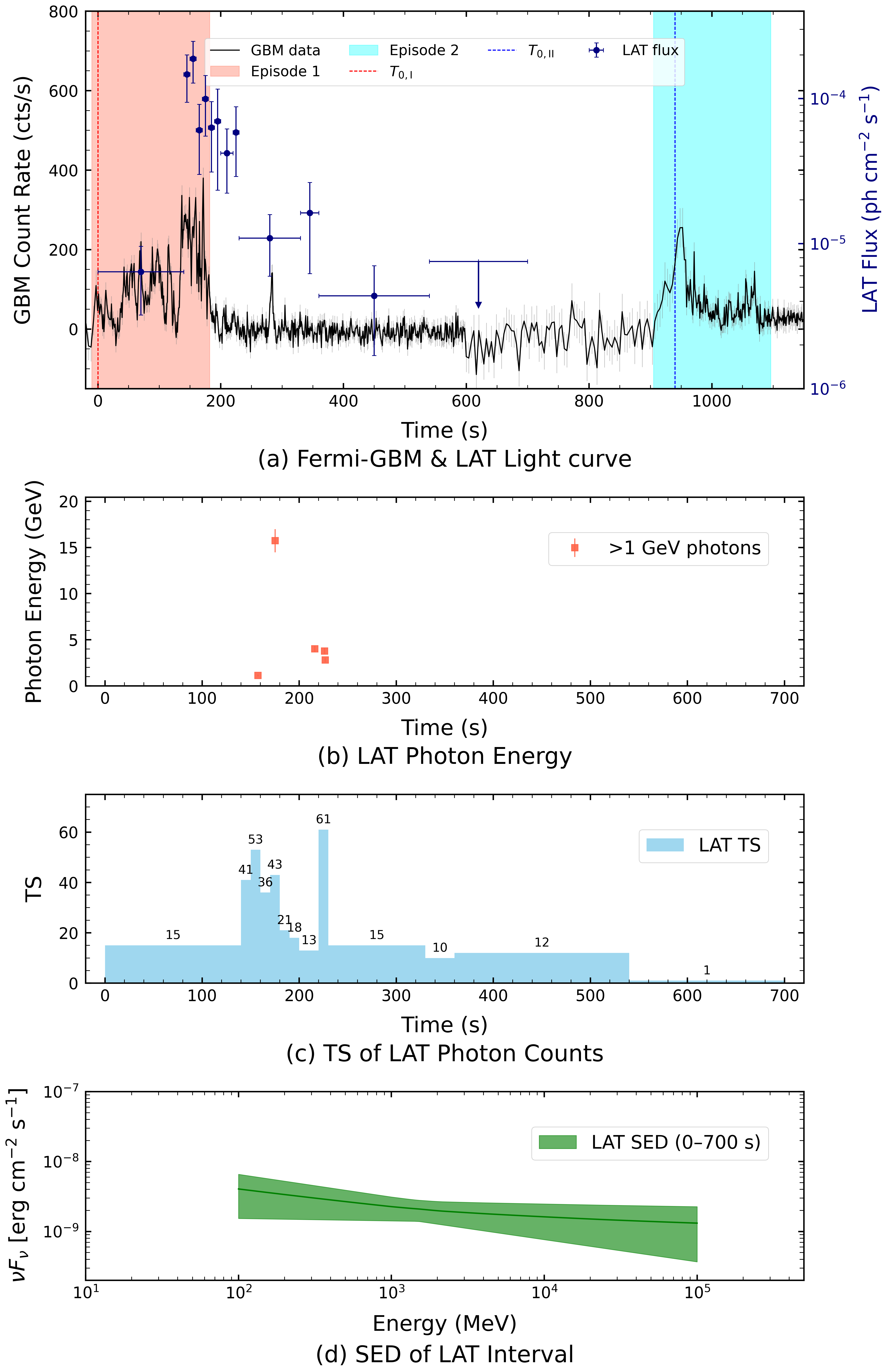} \\
    \small (a) Fermi-GBM \& LAT Light curve \& Spectral Features \\
    \vspace{10pt} 
    
    \includegraphics[height=0.3\textwidth, width=0.4\textwidth]{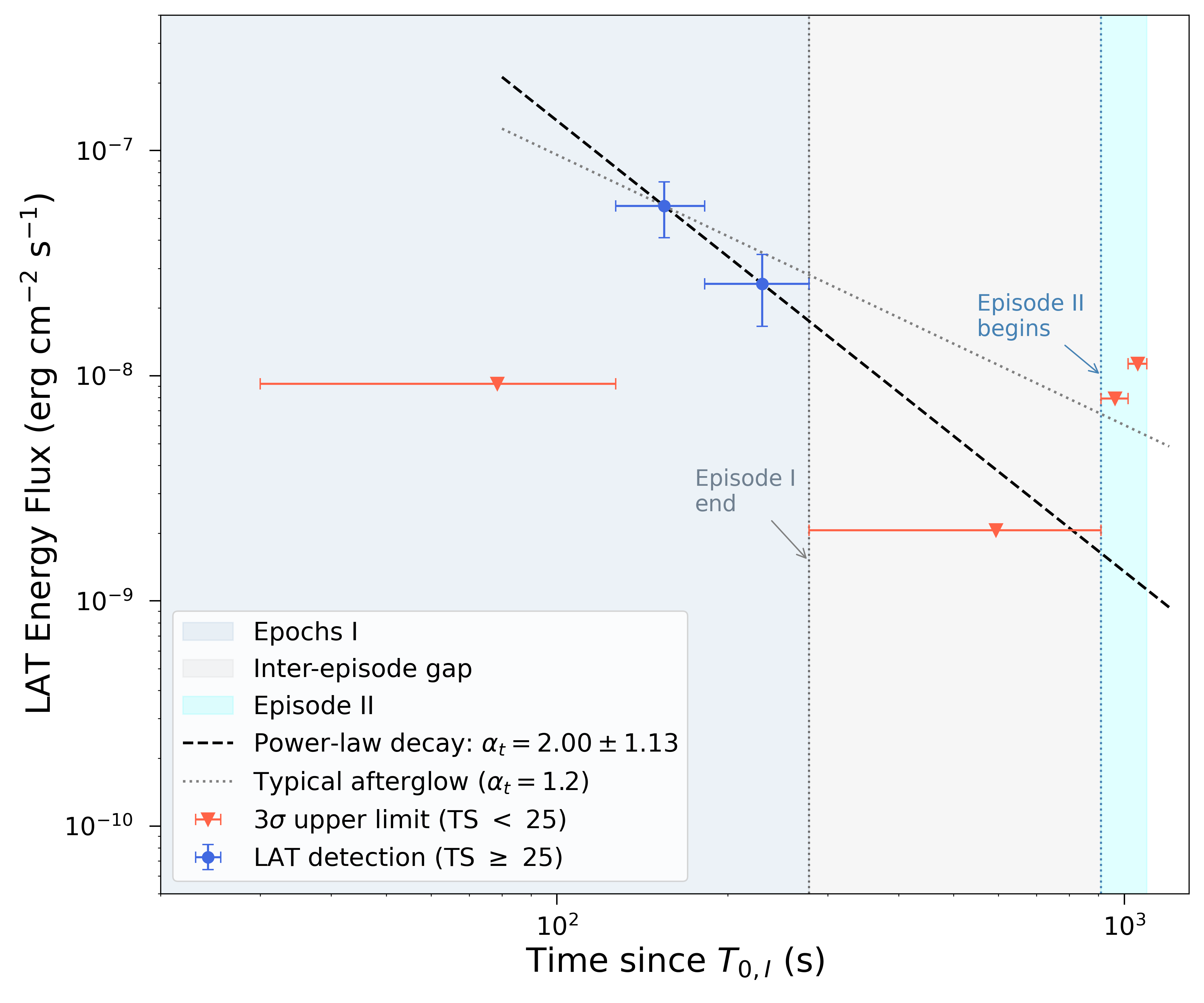} \\
    \small (b) LAT Energy Flux Light curve \\
    \vspace{10pt} 
    \caption{Fermi-LAT analysis of GRB~220627A ($0$--$700$~s from $T_{0_{(\rm I)}}$). (a) From top: LAT photon flux light curve) (blue) (100~MeV--300~GeV) on top of GBM net photon count rate (black), GeV photon counts (red), epoch-by-epoch TS values (sky blue histogram), and time-integrated spectral energy distribution fitted with a power law (green butterfly plot). (b) Energy flux light curve; filled circles (blue) indicate detections ($TS \geq 25$) and downward triangles (red) indicate $3\sigma$ upper limits ($TS < 25$). A power-law fit to the detected epochs (solid line) yields a temporal decay index $\alpha_t = 2.0 \pm 1.1$, steeper than the canonical afterglow slope $\alpha_t \sim 1.2$ (grey dashed; \citealt{Zhang2006, Nousek2006}), consistent with prompt-phase internal dissipation rather than forward-shock onset.}
    \label{fig:LAT_220627A}
\end{figure}

To investigate the high-energy regime, we constructed the Fermi-LAT light curve and SED for GRB~220627A from $0-700$~s using a standard unbinned likelihood analysis with TRANSIENT-class events in the 100~MeV-300~GeV range (\cite{Atwood2009}). 
We distributed the MeV and GeV LAT photons and highlighted the five $>1$~GeV photons in Figure~\ref{fig:LAT_220627A}.
The highest-energy photon, with energy 15.7 GeV, is observed at $T_0+175$~s, well within the prompt emission phase of episode--I.
The LAT emission is detected with a test statistic TS = 192 $(\sim 14\sigma)$ (\cite{Mattox1996}) and a time-averaged photon flux of $1.69~(\pm~0.26)\times 10^{-5}$~ph/cm$^{2}$/s within 700~s. 
A single PL model with spectral index $\lambda_{\mathrm{LAT}}~=~-2.18~\pm~0.15$ well describes the LAT spectrum, which is steeper than the GBM prompt indices ($\alpha~\sim~-0.7$ to $-1.1$), indicating spectral softening at GeV energies.

To investigate the epoch-by-epoch LAT behaviour within episode--I, we report the per-epoch TS values using the same time boundaries as the GBM flux-variability analysis. 
The LAT hardness-ratio test showed no significant trend for both episodes, as there were no photons in episode--II.
No LAT photons above threshold ($p > 0.95$) are associated with Epoch 1A (--10 to 30\,s).
During Epoch 1B (30--127\,s), the LAT yields TS = 5, consistent with no significant detection despite ongoing GBM flaring activity. 

The highest LAT significance occurs in Epoch 1C (127--182\,s; TS = 161, energy flux $= 5.68\times10^{-8}$\,erg\,cm$^{-2}$\,s$^{-1}$), coinciding with the brightest GBM flaring and the highest-energy 15.7\,GeV photon detected at $T_{0_1} + 175$\,s. 
A significant LAT detection persists into the early quiescent Epoch 1D (182--278\,s; TS = 83, energy flux $= 2.56\times10^{-8}$\,erg\,cm$^{-2}$\,s$^{-1}$), indicating that GeV emission does not cease instantaneously with the GBM flares but fades over $\sim200$\,s.
The full inter-episode gap (290--910\,s) shows no significant detection (TS = 24, marginally below threshold), and episode--II yields TS = 0 for both epochs~2A and 2B despite favourable observing conditions ($27^\circ$ boresight, no SAA passage), confirming that GeV emission is tied to the internal prompt dissipation of episode--I rather than to the development of an external forward shock \citep{Kumar2012, Beniamini2016}.

To characterise the LAT flux evolution across episode~I epochs, Figure~\ref{fig:LAT_220627A} shows the energy flux light curve with epoch-by-epoch TS values from the unbinned likelihood analysis. 
Significant LAT detections are obtained only during epochs~1C (127--182\,s; $TS = 161$, energy flux $= 5.68 \pm 1.58 \times 10^{-8}$~erg~cm$^{-2}$~s$^{-1}$) and 1D (182--278\,s; $TS = 83$, energy flux $= 2.56 \pm 0.90 \times 10^{-8}$~erg~cm$^{-2}$~s$^{-1}$), coinciding with the brightest GBM flaring (epoch~1C) and its decline (epoch~1D). 

Epoch~1B (30--127\,s) and all post-episode-I intervals yield only upper limits. A power-law fit to the two detections yields a temporal decay index $\alpha_t = 2.0 \pm 1.1$, steeper than the canonical afterglow slope ($\alpha_t \sim 1.2$; 


\cite{Zhang2006, Nousek2006}), consistent with prompt-phase decay rather than a forward-shock onset.

The LAT photon distribution (Figure~\ref{fig:LAT_220627A}, top panel) shows that the bulk of GeV emission falls within epochs~1B and~1C ($30-182$\,s), coinciding with the brightest GBM flaring activity and the hardest CPL indices ($\alpha\sim-0.44$ to $-0.72$).
The LAT spectral index ($\lambda_{\rm LAT}=-2.18\pm0.15$) is steeper than all GBM prompt indices ($\alpha\sim-0.7$ to $-1.1$) but does not show a statistically significant change across epochs, given the low photon statistics; we note this as a caveat and 
include the per-epoch LAT photon counts in Table~\ref{tab:GRB_TRS} for completeness.


Notably, no LAT photons are detected during episode--II $(T_0~+~900~-~1100$~s) despite favourable observing conditions ($27^\circ$~boresight angle, no SAA passage), indicating that the GeV emission is confined to the prompt emission (Figure~\ref{fig:LAT_220627A}) rather than extending into a smooth afterglow phase. 
\section{Discussion}
\label{sec:Discussion}

GRB~220627A is a transient event presenting two possible interpretations: an ul-GRB or a gravitationally lensed normal lGRB. Its prompt emission lasting $\geq10^3$~s, complex multi-peaked light curves spanning two distinct episodes separated by $\sim600$~s, and extreme radiative output matching population studies of ul-GRBs (\cite{Gendre2013, Levan2013}) must first be examined against the lensing hypothesis before any progenitor constraints can be drawn.


With lensing conclusively ruled out (\cite{Huang2022}; \ref{sec:AppendixA}), three progenitor scenarios compete to explain such ultra-long transient events: BSG collapsars, rapidly rotating newborn magnetars, and WD-TDEs (\cite{Gendre2013, Kashiyama2013, Perna2018}). 
We constrain these using Fermi-GBM time-resolved spectroscopy, flux variability, time-integrated spectral modelling, and Fermi-LAT emission, and compare throughout against ul-GRB progenitor templates: GRB~091024A (magnetar candidate) and GRB~130925A (BSG collapsar).

The three ul-GRB models produce observationally and theoretically distinguishable prompt emission light curves. BSG collapsars yield structured, multi-peaked bright $\gamma-$ray light curves lasting over $\sim 10^3-10^4$~s, accompanied by a gradual descent tracking the decay of the central engine luminosity, powered by episodic accretion through the extended Hydrogen envelope (\cite{Gendre2013, Perna2018}). Unlike BSG collapsars, the magnetar spin-down model produces a smooth plateau phase in the light curve governed by the spin-down luminosity, $L_\mathrm{sd}(t)~\propto~(1~+~t/\tau_\mathrm{sd})^{-2}$, terminating in a steep decay once the neutron star collapses, with total energy $E_\mathrm{rot} \lesssim 10^{53}$~erg (\cite{Zhang2001, Metzger2007, Lue2014, Fu2024}).

Furthermore, WD-TDE light curves produce long-lasting, highly variable X-ray emission driven by the $\dot{M}_\mathrm{fb} \propto t^{-5/3}$ fallback rate of disrupted white dwarf debris, over $\sim 10^4$~s, with characteristically soft prompt spectra ($E_p \lesssim 100$~keV) (\cite{Levan2013, Ioka2016}). Lastly, GRB~220627A's time-integrated structured, multi-peaked light curve, hard $E_p \sim 249$~keV, and $E_\mathrm{iso} = 1.6 \times 10^{54}$~erg exceeding magnetar spin-down limits by the order of 10, rule out both the magnetar and WD-TDE templates and favour the BSG collapsar scenario, Table~\ref{tab:multi-model}.


Time-resolved CPL spectroscopy reveals a hard-to-soft spectral evolution across both episodes: episode--I traces slow-cooling synchrotron emission ($\alpha~\sim~-0.44~\to~-1.07; E_p~\sim~206~\to~368$~keV, near the ``synchrotron line of death'' $\alpha\sim-0.67$), while episode--II shows systematic softening ($\alpha\sim-1.32\to-1.18$) with declining $E_p$, characteristic of fast-cooling dissipation ($\alpha\sim-1.5$; \cite{Piran2004, Uhm2014}; Figure~\ref{fig:Lightcurve_TRS}, Table~\ref{tab:GRB_TRS}). 
Comparing against established ul-GRB templates, GRB~220627A and the magnetar candidate GRB~091024A share a qualitatively similar hard-to-soft evolution (091024A: $\alpha_I\approx-0.74\to-0.79$, before softening in episode--II ($\alpha_{II}\approx-1.32$); Table~\ref{tab:GRB_TRS}), while the BSG prototype GRB~130925A instead exhibits a soft-to-hard trend within episode--I ($\alpha_I\approx-1.9\to-0.68$) before softening again in episode--II ($\alpha_{II}\approx-1.34$) (\cite{Greiner2014, Piro2014, Ghosh2024, Ror2024}). 

As established in the section~\ref{sec:Time-Resolved Spectroscopy},
this episode-resolved pattern closely matches the magnetar candidate GRB~091024A, while the BSG prototype GRB~130925A exhibits the reverse soft-to-hard trend in episode--I (\cite{Greiner2014, Piro2014, Ghosh2024, Ror2024}; Figure~\ref{fig:Lightcurve_GRB}(b)). However, while GRB~220627A's low-energy spectral index tracks GRB~091024A across both episodes, its episode--II spectral curvature is more consistent with GRB~130925A, preventing either progenitor model from uniquely explaining the spectral evolution and necessitating further diagnostics to resolve it.

Flux variability analysis shown in Table~\ref{tab:flux_variability}, using MVT values spanning $2.01 \pm 0.90$~s to $7.51 \pm 0.42$~s, places emission radii $R\sim10^{14}-10^{15}$~cm within the relativistic jet, precisely where electron cooling efficiency increases and magnetic field and particle injection evolve (\cite{Uhm2014, Derishev2024}). The CCF lag (45~s), obtained from a flare variations study via Fourier transform, further traces fallback accretion timescales or jet propagation through density variations in an extended stellar envelope (\cite{Nakauchi2013, Lazzati2017, Perna2018}). 

Critically, our time-resolved spectral index evolution complements \cite{Wet2023}'s time-integrated afterglow parameters: electron spectral index ($p~\approx~2.0$) and a circumburst medium density ($n~\approx~20~\rm{cm}^{-3}$), demanding prolonged central engine variability far beyond standard forward-shock predictions. Using coasting Lorentz factors $\Gamma_c\sim3{-}4$ typical for BSG jets (\cite{Goldstein2017}), the photospheric radius $R_\mathrm{ph}~\sim~10^{14}$~cm derived from eq.~\ref{eq:rph} matches our MVT-constrained dissipation radii (Table~\ref{tab:flux_variability}). This $R_\mathrm{ph}~=~R_\mathrm{MVT}$ equality is consistent with the BSG geometry, since magnetars lack extended envelopes for cocoon formation while WD-TDEs predict much softer thermal emission ($kT\lesssim10$ keV) (Table~\ref{tab:multi-model}) (\cite{Ioka2016, Lazzati2017}).

For the time-integrated prompt emission of GRB~220627A, we fitted the GBM spectra of both episodes with CPL, BB, and CPL+BB models (Table~\ref{tab:GBM_full_spectral}). Episode--I ($-6$~s, 181.6~s) is characterised by $E_p = 317.21$~keV and $\alpha_I = -0.70$, lies close to the slow-cooling synchrotron "line of death" ($\alpha~\sim~-0.67$), while episode--II ($-46$~s, 140~s) shows $E_p~=~231.35$~keV and $\alpha_{II}~=~-1.29$, approaches the fast-cooling limit ($\alpha\sim-1.5$). The shift in the spectral indices, crossing the canonical synchrotron limits, indicates a transition from slow to fast radiative cooling between episodes (\cite{Preece1998, Zhang2016}).

The total integrated $E_\mathrm{iso}~=~1.56~(\pm~0.14) \times 10^{54}$~erg at $z = 3.084$ exceeds GRB~091024A ($E_\mathrm{iso}~=~1.9~\times~10^{53}$~erg, $\sim 8~\times$ lower) and GRB~130925A ($E_\mathrm{iso}=1.15 \times 10^{52}$~erg, $\sim~13~\times$ lower), surpassing magnetar spin-down limits ($\sim~10^{53}$~erg). These values, alongside the CPL-derived total $E_p~=~248.90~\pm~8.39$~keV, multi-peaked morphology, and $T_{90}>1100$~s, align with BSG collapsars such as GRB~101225A, GRB~111209A/ SN~2011kl, and GRB~130925A (\cite{Gendre2013, Levan2013, Greiner2015}), while the extreme energetics rule out WD-TDE energy budgets and the structured variability is incompatible with the smooth magnetar spin-down luminosity profile (\cite{Krolik2011, Nuessle2024, Ror2024}). In episode--resolved $E_p-L_\mathrm{iso}$ space, GRB~220627A occupies classic BSG territory alongside GRB~130925A, favouring extended envelope, jet-cocoon interaction over central engine spin-down mechanism (\cite{Virgili2013, Evans2014}).

Fermi-LAT detection provides a decisive constraint on the emission mechanism. Significant GeV emission is detected exclusively during episode--I ($TS~=~192,~\sim~14~\sigma;~\lambda_{\rm LAT} = -2.18 \pm 0.15$), with the highest-energy photon of 15.7~GeV arriving at $T_0+175$~s during the main flaring phase, and no LAT emission detected during episode--II despite favourable observing conditions ($27^\circ$ boresight angle, no SAA passage). This precise temporal correlation with the GBM prompt flares, rather than the smooth rise and PL decay expected from external forward-shock afterglows, immediately favours an internal dissipation origin (\cite{Kumar2012, Beniamini2016}). 

The 15.7~GeV photon ($T_0+175$~s) exceeds external shock synchrotron limits: 
$$E_{\rm syn,~max} \sim 50\frac{\Gamma}{1+z}~\rm MeV \quad (\Gamma\sim300{-}1000)$$
requires internal dissipation with enhanced magnetic fields or synchrotron self-Compton (SSC) upscattering of prompt MeV photons within the jet (\cite{Wang2013, Beniamini2013, Ravasio2024}).
The LAT spectral index $\lambda_{\rm LAT} = -2.18$, steeper than the GBM prompt indices ($\alpha \sim -0.7$ to $-1.1$), matches the fast-cooling synchrotron high-energy tail or SSC emission above the synchrotron peak, as established for LAT-detected GRBs~080916C and 190114C (\cite{Daigne1998, Kumar2008, Bosnjak2009, Abdo2009, Mohnani2026}). 
The absence of GeV emission in Episode–II, despite comparable GBM characteristics, indicates that the two episodes exhibit different high-energy emission. 
No emission is detected above the GBM background in the $\sim~600$~s gap, so it is hard to confirm whether the engine was dormant or emitting below our sensitivity (\cite{Bosnjak2009, Hascoet2012}).

The combined GBM and LAT light curve (Figure~\ref{fig:LAT_220627A}) allows a direct test of whether an afterglow is initiated at the end of episode--I or at the conclusion of the LAT-detected emission. 
Multiple independent lines of evidence argue against afterglow onset at either boundary. 
Following the end of episode--I ($\sim$182~s), the GBM count rate becomes background-dominated and remains statistically consistent throughout the entire $\sim$600~s inter-episode gap, exhibiting no trace of the smooth, monotonically decaying power-law profile ($F\propto t^{-\alpha}$, $\alpha_t\sim1$--1.5) expected from an established external forward shock (\citealt{Zhang2006, Nousek2006}). 
The subsequent re-brightening into episode--II at~$\sim$905~s, reaching peak count rates comparable to episode--I ($\bar{x}_{2\rm A}\approx276$~ph~cm$^{-2}$~s$^{-1}$; Table~\ref{tab:flux_variability}), requires a reactivated central engine: an adiabatically decelerating external shock cannot re-energise to these levels without sustained energy injection, and the rapid variability amplitudes ($F_\mathrm{var}=0.47$--0.59) and MVTs of 4--8~s in episode--II are irreconcilable with external-shock emission (\citealt{Derishev2024}). 
This distinguishes GRB~220627A from bursts in which the GBM light curve explicitly traces a declining power law coincident with the LAT decay after the prompt pulses, as found for GRB~200524A ($\alpha_t \approx -1.79$; \citealt{Ghosh2026}), which is associated with the early afterglow onset.

Turning to the Fermi-LAT light curve, GeV emission is strictly contemporaneous with GBM episode--I prompt flares: TS values peak during the brightest GBM flaring (epochs~1C--1D, 127--278~s) and drop immediately to non-detection simultaneously with the dormant engine activity, with episode--II yielding $TS~=~0$ despite a $27^\circ$ boresight and no SAA passage. 
An external forward-shock afterglow would instead produce GeV emission that peaks \textit{after} the prompt episode and persists smoothly through the inter-episode gap and episode--II (\citealt{Kumar2012, Beniamini2016}). 
The LAT temporal decay index $\alpha_t=2.0\pm1.1$ is steeper than the canonical afterglow slope ($\alpha_t\sim1.2$; \citealt{Zhang2006, Nousek2006}), consistent with renewed central engine activity 
rather than forward-shock onset. 
Furthermore, the 15.7~GeV photon at $T_0+175$~s exceeds the external synchrotron burn-off limit, demanding an internal dissipation origin (\citealt{Wang2013, Beniamini2013}), and multi-wavelength afterglow modelling detects the external forward shock only at $\sim$1.5~d post-burst (\citealt{Wet2023}), confirmed by radio monitoring (\citealt{Leung2026}). 
Collectively, the GBM+LAT overlay and the multi-wavelength afterglow chronology confirm that episode--II is a second genuine prompt-emission episode driven by central engine reactivation, with no evidence for external shock initiation at either episode boundary.

Synthesising the time-integrated results, four independent diagnostics break the progenitor degeneracy toward BSG: (1) $E_\mathrm{iso}=1.6 \times 10^{54}$~erg  ($\sim 8\times$ GRB~091024A), far exceeding magnetar spin-down limits; (2) the break period of 600~s requires fallback accretion through an extended envelope, absent in magnetar spin-down models; (3) a possible thermal signature from CPL+BB fits during flares ($kT~\sim~22~-~31$~keV, time-resolved; $kT~\approx~30~-~31$~keV, time-integrated); and (4) the radio-inferred cocoon photosphere $R_\mathrm{ph} \approx 1.3 \times 10^{13} (M/10M_\odot)^{-1/2}$~cm (\cite{Leung2026}), consistent with jet propagation through an extended hydrogen-rich BSG envelope. Magnetars predict smooth X-ray plateaus with $E_\mathrm{iso}~\ll~10^{54}$~erg and no structured variability; WD-TDEs demand $E_p~\lesssim~100$~keV in low star-formation environments (\cite{Krolik2011, Gendre2013, Virgili2013, Gompertz2017}). 
Radio-inferred cocoon energetics yield a massive stellar envelope ($\sim100~R_\odot$; \cite{Leung2026}), further supporting BSG collapse, with our time-resolved $E_p$ decline tracing progressive envelope cooling, not captured by their wide-ejecta afterglow model.

This hybrid thermal/non-thermal 
spectral structure is consistent with BSG collapsar origin through jet-cocoon interaction. A possible thermal component with $kT \sim 22-31$~keV is suggested during major flaring intervals in both episodes, persisting even when NaI K-edge channels ($kT \sim 31-35$~keV) are excluded, which rules out a straightforward instrumental origin (\ref{sec:AppendixB}; Table~\ref{tab:GBM_full_spectral}, Figure~\ref{fig:GBM_thermal}). The time-integrated CPL+BB analysis yields $kT \sim 30-31$~keV with a modest improvement in fit quality (lower PGStat/DOF) compared to CPL-only. CPL+BB fit reveals systematic evolution from episode--I ($\alpha = -0.82$, $kT \sim 30.41$~keV) to episode--II ($\alpha = -1.60$, $kT \sim 31.34$~keV), tracking the non-thermal hardening/softening pattern and demonstrating that declining fallback accretion modulates both cocoon temperature and dissipation efficiency.

Theoretical cocoon temperatures $T_{\rm cocoon}~\sim~10^8{-}10^9$~K; ($kT \sim 10-100$~keV; \cite{Lazzati2017}) encompass the observed range (Table~\ref{tab:thermal_grbs}). As the cocoon expands and becomes optically thin, thermal seed photons decouple at $R_\mathrm{ph}\sim10^{12}$~cm (\cite{Rees2005, PeEr2012, Zhang2018}) and propagate outward to our MVT-derived dissipation radii $R_\mathrm{diss}\sim10^{14}-10^{15}$~cm, where Inverse Compton upscattering by PL electrons produces the observed MeV CPL curvature superposed on the thermal excess (\cite{PeEr2006, Giannios2008, Beloborodov2010}). This explains GRB~220627A's quasi-stable episode--I $kT$ (uniform inner-envelope conditions), episode--II spectral cooling through outer envelope layers, and the suppressed BB contribution during quiescent periods when the photospheric photon supply is reduced.

Such photospheric emission finds analogous thermal excesses in confirmed BSG ul-GRBs: GRB~130925A ($kT \sim 10-50$~keV) and GRB~111209A/ SN~2011kl (\cite{Piro2014, Greiner2015, Lazzati2017}; Table~\ref{tab:thermal_grbs}).  
The hybrid thermal/non-thermal spectral shape, stratified jet structure (thermal cocoon core + magnetised dissipative sheath), and $R_\mathrm{ph}\sim10^{12}$~--~$R_\mathrm{diss}\sim10^{14-15}$~cm consistency represent established diagnostics of massive-star jet-cocoon interaction, increasingly recognised in ul-GRB/ lGRB studies (\cite{Ryde2010, Axelsson2012, Ito2013, Nakauchi2013, Evans2014, Piro2014, Lazzati2017, Ravasio2024}). 

More broadly, strong photospheric thermal components alongside non-thermal emission have also been identified in energetic classical long GRBs such as GRB~230204B ($E_\mathrm{iso} \sim 2.2 \times 10^{54}$~erg), whose wind-medium afterglow independently confirms a massive star progenitor origin (\cite{Swain2026}), reinforcing the interpretation that the hybrid thermal/non-thermal spectral structure of GRB~220627A reflects jet-cocoon photospheric emission in a collapsar environment.


Thus, GRB~220627A emerges as the most luminous, structurally complex BSG collapsar detected, with time-resolved spectroscopy revealing a magnetar-like spectral degeneracy, resolved by decisive energetic, radio-cocoon, and possible photospheric signatures, thereby demanding fallback accretion through an extended hydrogen envelope.

\section{Conclusion}
\label{sec:Conclusion}

This work presents a detailed prompt-emission study of the ul-GRB~220627A using Fermi-GBM and Fermi-LAT data, combining time-resolved spectroscopy, flux-variability analysis, and high-energy $\gamma-$ray characterisation. 
The main conclusions are:

\begin{itemize}
    \item The two-episode structure, total prompt duration $T_{90}~\geq~10^3$~s, and extreme energetics $E_{iso}~\simeq~1.6~\times~10^{54}$~erg place GRB~220627A among the most luminous ul-GRBs known, incompatible with being a simple gravitationally lensed duplicate of a shorter burst.

    \item Time-integrated and time-resolved GBM spectra are well described by a CPL model, with episode--I showing $\alpha~=~-0.7$ and episode--II $\alpha~=~-1.3$, consistent with a transition from slow- to fast-cooling synchrotron emission in an internal jet dissipation region (\cite{Preece1998, Uhm2014}).

    \item Thermal components ($kT~\sim~22-31$~keV) during flares, analysed via CPL+BB fits excluding NaI K-edge effects (\ref{sec:AppendixB}, Figure~\ref{fig:GBM_thermal}) and time-integrated study of hybrid model yield ($kT~\sim~30-31$~keV) Table~\ref{tab:GRB_TRS}, indicate jet-cocoon/photospheric emission characteristic of BSG collapsars, supporting a hybrid thermal and non-thermal prompt spectrum (\cite{PeEr2006, Lazzati2017}).

    \item LAT detects significant GeV emission only during the episode--I, with a $15.7$~GeV photon at $T_0+175$~s and a PL spectrum with $\lambda_{LAT}~=~-2.18$, temporally coincident with GBM flares and inconsistent with a smooth external-shock afterglow. Internal synchrotron/SSC processes within the jet provide a rational explanation for the observed GeV properties (\cite{Bosnjak2009, Beniamini2015}).

    \item GRB 220627A, the most luminous ($E_{\rm iso}~=~1.6~\times~10^{54}$~erg), structurally complex ul-GRB, exhibits time-resolved spectral evolution matching magnetar GRB 091024A alongside BSG compatible time-integrated high $E_{iso}$ and $L_{iso}$, internal emission radii, transient photospheric signatures, and internal GeV emission, establishing previously unresolved progenitor degeneracy at temporal resolution (\cite{Levan2013, Perna2018, Wet2023}).
\end{itemize}

Future studies of the ul-GRB population combining higher-cadence time-resolved spectroscopy (sub-second HOP across the full energy range) with polarisation measurements will resolve progenitor degeneracy, quantify the jet-cocoon coupling efficiency, and distinguish between fallback accretion and magnetar spin-down timescales across diverse environments.


\appendix

\appendix
\section{Correlation Analysis: Ruling Out Gravitational Lensing}
\label{sec:AppendixA}

Gravitational lensing was conclusively ruled out at $5.1\sigma$ by \citep{Huang2022} based on asymmetric LAT emission. 
For completeness, we present CCF and PSD analyses that independently confirm episodes via correlated but time-delayed GBM activity (Table~\ref{tab:CCF_results}) and an evolving PSD spectral index (Table~\ref{tab:PSD_results}), consistent with fallback accretion rather than magnified copies.
%

GRB~220627A: The peak CCF~$=~0.511$ at spectral lag of $+~44.9$~s (SNR~$=~2.8$) shows episode--II follows episode--I with measurable delay, consistent with recurrent accretion episodes rather than magnified copies (\cite{Nakauchi2013}). 

GRB~091024A: Weak CCF $= 0.248$ at 1~s lag suggests loosely coupled episodes, consistent with magnetar spin-down models where rapid re-brightening occurs (\cite{Gendre2013, Virgili2013}).

\begin{table}[H]
\centering
\begin{adjustbox}{width=0.45\textwidth, center}
\renewcommand{\arraystretch}{1.5} 
\Large
\begin{tabular}{lcccc}
    \hline
    \textbf{GRB} & \textbf{Peak CCF} & \textbf{Lag (s)} & \textbf{SNR} & \textbf{Interpretation} \\
    \hline
    GRB 220627A & 0.511 & +44.9 & 2.8 & Correlated, delayed \\
    GRB 091024A & 0.248 & +1.0  & 3.1 & Weak correlation \\
    GRB 130925A & 0.229 & -84.7 & 7.0 & Anti-correlation (distinct episodes) \\
    \hline
\end{tabular}
\end{adjustbox}
\caption{Peak CCF values along with their associated SNR and spectral lag results between episode--I and II for all three ul-GRBs is defined along with the interpretations.}
\label{tab:CCF_results}
\end{table}

GRB~130925A: Anti-correlation (CCF~$=~0.229$ at spectral lag of -84.7~s) shows distinguishing physical processes between episodes, consistent with jet propagation through stratified envelope density (\cite{Evans2014, Piro2014}).

Our time-resolved analysis reveals a strong flux-hardness correlation (\cite{Nousek2006}), consistent with synchrotron cooling from episodic cocoon shocks (\cite{Leung2026}).

FFT-estimated PSDs are used to measure frequency-domain variability (\cite{Welch1967, Klis1989}). 

GRB~220627A: PSD decreases from $\beta_{\rm I}~=~-1.24$ (pink noise) to $\beta_{\rm II}~=~-1.51$ (Brownian noise), indicating a transition toward lower-frequency variability. 
This reflects reduced rapid dissipation in episode--II (lower accretion rate, larger emission region), as predicted by fallback accretion models (\cite{Nakauchi2013}).

GRB~091024A: Stable PSD ($\Delta\beta~=~+0.04$) consistent with magnetar spin-down producing similar variability across episodes (\cite{Gendre2013}).

GRB~130925A: Dramatic steepening ($\Delta\beta~=~-2.61$) suggests a substantial change in jet structure or envelope interaction (\cite{Evans2014}).


\begin{table}[H]
\centering
\begin{adjustbox}{width=0.45\textwidth, center}
\renewcommand{\arraystretch}{1.3} 
\Large
\begin{tabular}{lcccc}
\hline
\textbf{GRB} & \textbf{Episode I $\beta$} & \textbf{Episode II $\beta$} & 
\textbf{$\Delta\beta$} & \textbf{PSD} \\
 &  &  &  & \textbf{Evolution} \\
\hline
GRB 220627A & -1.24 & -1.51 & -0.27 & Steeper (redder) \\
GRB 091024A & -1.69 & -1.65 & +0.04 & Stable \\
GRB 130925A & -0.55 & -3.16 & -2.61 & Much steeper \\
\hline
\end{tabular}
\end{adjustbox}
\caption{PSD PL indices ($\beta$) for episode--I and II are defined for all three ul-GRBs along their evolution behaviour.}
\label{tab:PSD_results}
\end{table}

Extending beyond \cite{Huang2022}'s ruling out of lensing via LAT asymmetry, our GBM time-resolved CCF lags and PSD confirm independent episodes, with spectral indices matching GRB~091024A/130925A patterns but sharper thermal transients indicating unique GRB~220627A envelope interaction.
This is distinct from magnetar models (stable PSD, rapid re-brightening) and WD-TDE (softer spectra, lower energetics), favouring BSG collapsar origin (\cite{Levan2013, Ioka2016}).

\section{Thermal Study} 
\label{sec:AppendixB}

\begin{figure}[H]
\centering
    \includegraphics[width=0.4\textwidth]{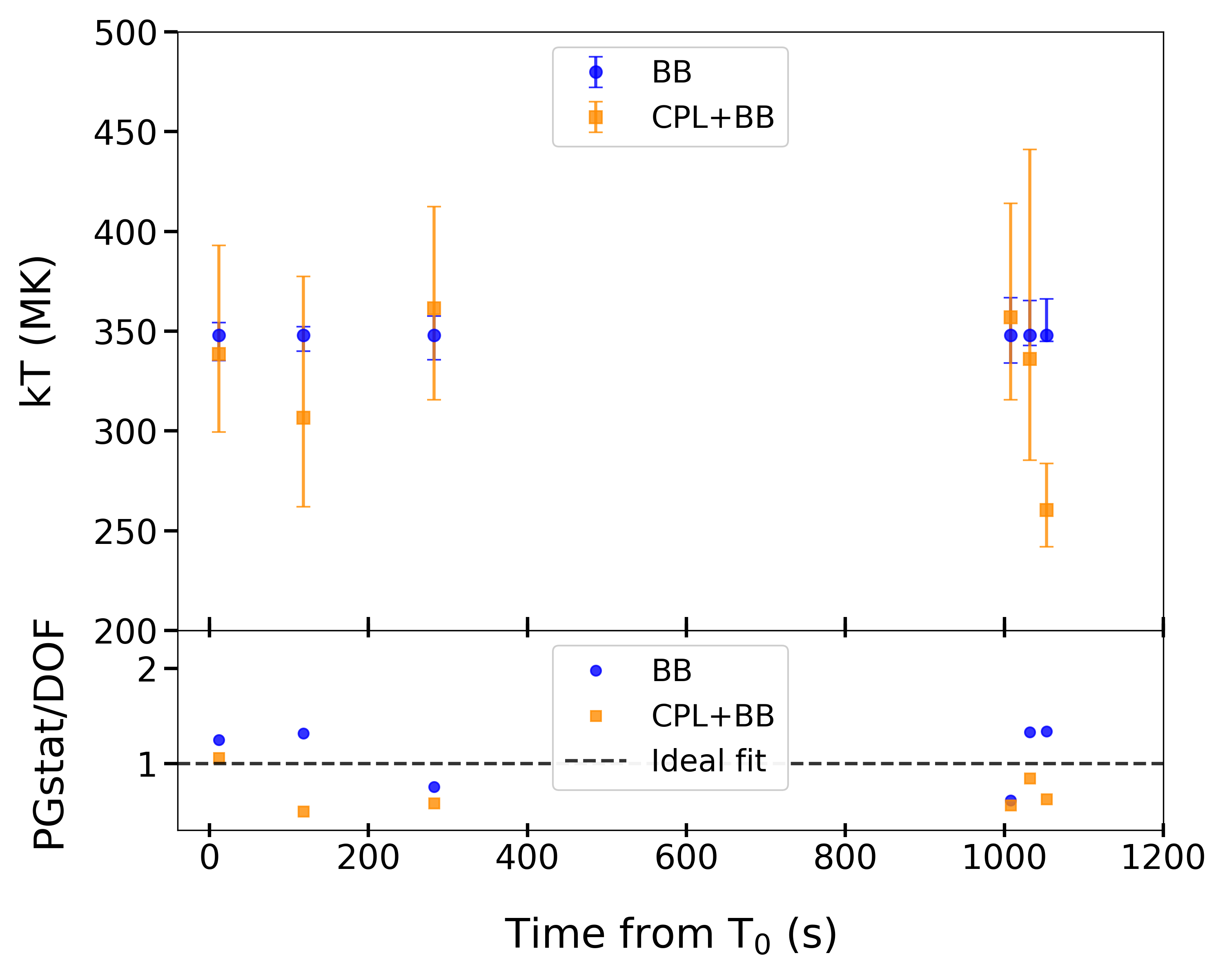} \\
  \caption{Black-body temperature evolution $kT(t)$ from BB only and CPL+BB fits (31-35 keV excluded), overlaid with PGStat/DOF. For CPL+BB fits, Episode--I: $kT \to 28-32$~keV (stable high-T photosphere); Episode--II: $kT \to 22-31$~keV (modest cooling). A horizontal dashed line at PGStat/DOF$~\simeq~1$ indicates an ideal fit for statistics.}
  \label{fig:GBM_thermal}
\end{figure}

Transient thermal signatures were detected in selective bright flaring intervals of both episodes using the BB model.
With only BB fit, the thermal component is identified, with many bins converging on the analysis prior upper limit $kT~=~29.99$~keV, and minor asymmetric uncertainties are well below the NaI K-edge (33.17~keV), as shown in Figure~\ref{fig:GBM_thermal}. 
The subsequent CPL+BB analysis, performed with 31-35 keV channels explicitly excluded to eliminate any possible K-edge influence, refines the central values of BB temperatures to $kT~\simeq~22.4{-}31.2$~keV with PGStat/DOF $\simeq~0.5{-}1.2$. 
Since the edge-affected channels are excluded from the fit, the quoted positive uncertainties reflect only statistical variation within the unaffected spectral range. 

To accurately model both thermal and non-thermal spectral components, simultaneous CPL+BB fits were performed:
\begin{equation}
\begin{aligned}
\frac{dN}{dE} &= A_{\rm CPL} \left(\frac{E}{E_{\rm piv}}\right)^\lambda e^{-(2+\lambda) E/E_p} \\ 
&\quad + A_{\rm BB} \frac{E^2}{(e^{E/kT} - 1)}
\end{aligned}
\label{eq:cpl_bb}
\end{equation}
where $A_{\rm CPL}$ is normalisation flux value (ph/cm$^2$/s/keV) for CPL, $\lambda$ is CPL spectral index, $E_p$ is the SED peak energy (keV), $E_{piv}$ is the pivot energy (in keV) fixed to 100~keV for CPL model, and $A_{\rm BB}$ is normalisation flux value (ph/cm$^2$/s/keV) for BB, and $kT$ is temperature (keV) for BB model.

CPL+BB fits with 31-35 keV channels excluded yield:
\begin{itemize}
\item Improved CPL spectral indices with reduced uncertainties compared to CPL-only fits.
\item Possible BB temperatures constrained at $kT~\sim~22-31$~keV with PGStat/DOF~$\simeq~0.5-1.2$ and time-integrated thermal value $kT~\sim~30-31$~keV.
\item Consistent thermal amplitudes across multiple independent fitting trials, indicating stable solutions.
\end{itemize}

The improvement in fit quality (lower PGStat/DOF with CPL+BB versus CPL-only) provides statistical evidence that both thermal $\&$ non-thermal radiation components are genuinely present in the observed spectrum, rather than one component being spurious.



Theoretical models predict the following cocoon characteristics:
\begin{itemize}
\item \textit{Temperature}: $T_{\rm cocoon}~\sim~10^8-10^9$~K, equivalent to $kT~\sim~10-100$~keV (\cite{Lazzati2017}).
\item \textit{Optical depth}: Initially optically thick (spectral lag $\gg 1$) due to high density and Thomson opacity.
\item \textit{Thermal time frame}: Cocoon extends the dynamic timescale of jet propagation (seconds to minutes, depending on the envelope density structure).
\end{itemize}

As the cocoon grows, it gradually becomes optically thin, releasing trapped thermal radiation. 
The photospheric radius, where optical depth $\tau~=~1$ (\cite{Rees2005, Zhang2018}), evolves as:
\begin{equation}
R_{\rm ph}(t) \approx 10^{12} \left(\frac{\Gamma_c}{4}\right)^2 \left(\frac{t}{1 \text{ s}}\right) \text{ cm},
\label{eq:rph}
\end{equation}
where $\Gamma_c \sim 3-4$ is the coasting Lorentz factor for BSG jets.


This mechanism explains:
\begin{itemize}
\item Transient thermal detections during flares (fresh photospheric supply).
\item Hard-to-soft evolution ($\alpha_{\rm I}~=~-0.70$ $\to \alpha_{\rm II}~=~-1.29$) from declining seed photon input.
\item CPL+BB spectral improvement (Table~\ref{tab:GRB_TRS}).
\end{itemize}


Figure~\ref{fig:GBM_thermal} shows the evolution of BB temperature $kT(t)$ from the \textit{CPL+BB} fits with 31-35~keV channels excluded. 
\textit{Episode--I} ($T_0<300$~s): $kT~\sim~26-31$~keV ($\Delta kT~\sim~3-6$~keV), with stable values indicating a persistent high-temperature photosphere/cocoon during peak jet activity.
\textit{Inter-episode gap} ($300<T<900$~s): background-dominated with no emission detected, so $kT$ is not constrained.
\textit{Episode--II} ($T>900$~s): $kT~\sim~22-31$~keV ($\Delta kT~\sim~2-9$~keV), with systematically lower central values than episode--I. 
While absolute uncertainties are larger, the fractional errors remain modest ($\Delta kT/kT \sim 10-15\%$), confirming thermal temperature determination even in the CPL+BB hybrid fit. 

Using equation~\ref{eq:rph} with $\Gamma_c\sim4$ and the characteristic soft-spectrum timescale $T_{\rm soft}\sim100$~s, which corresponds to photospheric cooling in the jet comoving frame at the photospheric radius $R_{\rm phos,\,c}~\simeq~10^{14}$~cm, we find thermal emission from shocked jet-cocoon material peaks before transitioning to non-thermal synchrotron, remarkably consistent with our photospheric radius estimate.
This photospheric radius estimate is remarkably consistent with the emission radius $R~\sim~10^{14}-10^{15}$~cm derived independently from MVT analysis in Table~\ref{tab:flux_variability}. 
This agreement provides significant cross-verification that the measured thermal signature originates in the cocoon/photospheric region rather than in more distant afterglow zones, reinforcing confidence in the jet-cocoon explanation.

\section*{Acknowledgements}
We are grateful to the Council of Scientific and Industrial Research (CSIR) for funding AG via a CSIR-JRF fellowship under grant 09/1022(13386)/2022-EMR-I. 
This research was funded in part by the National Science Center (NCN), Poland, grant number OPUS 2021/41/B/ST9/00757. 
The URJA research group in the IIT Indore department of Astronomy, Astrophysics, and Space Engineering provided computational resources.
For Open Access, the author has applied a CC-BY public copyright licence to any Author Accepted Manuscript (AAM) version arising from this submission.
\vspace{-1em}

\section*{Data Availability}
The analysed data underlying this work will be shared with the corresponding author(s) upon reasonable request.





\bibliography{ref}

\end{document}